\documentclass[12pt, a4paper]{article}
\usepackage{amsfonts, amsbsy, amsmath, amssymb, thmtools}
\usepackage[english]{babel}
\usepackage{bm}
\usepackage{booktabs}
\usepackage[final]{changes}
\usepackage{color, colortbl}
\usepackage{csquotes}
\usepackage{dsfont}
\usepackage{endnotes}
\usepackage{fancyhdr}
\usepackage{enumerate} 
\usepackage{fleqn}
\usepackage{float}
\usepackage[T1]{fontenc}
\usepackage{footnote}
\usepackage{graphicx}
\usepackage[style=apa]{biblatex}
\usepackage[utf8]{inputenc}
\usepackage{longtable}
\usepackage{lmodern}
\usepackage{lscape}
\usepackage[cal=cm]{mathalfa}
\usepackage{mathtools}
\usepackage{multirow, multicol}
\usepackage{rotating}
\usepackage{soul}
\usepackage{setspace}
\usepackage{tabularx, tabulary}
\usepackage{esvect}
\usepackage{geometry}
\usepackage{pdfpages}
\usepackage{hyperref, url}
\usepackage{verbatim}
\usepackage{caption}
\usepackage{adjustbox}
\usepackage{subcaption}
\usepackage{adjustbox}

\begin{document}

%%%%%%%%%%%
%% TITLE PAGE %%
%%%%%%%%%%%
	
\begin{center}
{\bf \Large Influencing a Polarized and Connected Legislature}\\
		
\vspace*{0.5cm}
		
{\large Ratul Das Chaudhury$^1$, C. Matthew Leister$^1$, Birendra Rai$^{1 \ast}$}\\
\end{center}
\bigskip
\bigskip
\begin{center}
ABSTRACT\\
\end{center}
	
\noindent When can an interest group exploit polarization between political parties to its advantage? Building upon Battaglini and Patacchini (2018), we study a model where an interest group credibly promises payments to legislators conditional on voting for its preferred policy. A legislator can be directly susceptible to other legislators and value voting like them. The overall pattern of inter-legislator susceptibility determines the relative influence of individual legislators, and therefore the relative influence of the parties. We show that high levels of ideological or affective polarization are more likely to benefit the interest group when the party ideologically aligned with the interest group is relatively more influential. However, ideological and affective polarization operate in different ways. The influence of legislators is independent of ideological polarization. In contrast, affective polarization effectively creates negative links between legislators across parties, and thus modifies the relative influence of individual legislators and parties.\\
	
	
\bigskip

\hrule
\bigskip
\noindent $^1$ Department of Economics, Monash University, Clayton, VIC 3800, Australia\\
\noindent $^*$ Corresponding Author: birendra.rai@monash.edu\\
\noindent {\it Acknowledgments.} We thank Marco Faravelli, Jun Xiao, Yves Zenou, and conference participants at the American Economic Association Meeting (2021) and the Econometric Society European Meeting (2021). We are particularly grateful to three anonymous referees and the editor for extensive suggestions. The usual disclaimer applies. 
	
\newpage
	
%%%%%%%%%%%
%% INTRODUCTION %%
%%%%%%%%%%%

\noindent \textbf{\Large 1. Introduction} \\
	
\noindent Scholars have long investigated whether material and informational exchange between special interest groups and legislators impacts legislative processes and outcomes (Groseclose and Snyder, 1996; Hall and Deardorff, 2006; Schnakenberg, 2017; Battaglini and Patacchini, 2018). Over the last two decades, several studies have documented a sharp increase in polarization between political elites in the United States (McCarty et al., 2006; Gentzkow et al., 2019; Iyengar et al., 2019). However, the impact of rising polarization between political parties on the ability of an interest group to influence legislation has not been systematically explored. The present paper aims to fill this gap.

It seems plausible that an interest group ultimately seeks a legislature that is sympathetic to its interests. The existing literature largely agrees that interest groups target investments toward some legislators to induce them to mobilize broader support for the interest group (Bauer et al., 1972; Shepsle, 1978; Denzau and Munger, 1986).\footnote{A related line of research emphasizes the role of issue-specific information and arguments that an interest group can provide (Hall and Deardorff, 2006; Schnakenberg, 2017).} A bill becomes a law through a complex multi-stage process often involving amendments to the original bill and compromises between legislators within and across parties on the form of the bill to be voted upon. Hence, the support an interest group seeks is not only limited to securing the final vote of a legislator on a particular bill. It also seeks accommodation of its preferences in the substantive content of every bill pertinent to its interests (Hall and Wayman, 1990; Ansolabehere and Snyder, 1998; Barber, 2016). 

Arguably, the extent to which a particular legislator can mobilize support in favor of the interest group will depend on, among other things, their capacity to influence other legislators. One source of such influence is the official power of a legislator due to their position in the legislature. For instance, several studies document the official power of legislators on committees charged with setting rules of legislative procedure, and those who can assign legislators to various committees (Romer and Snyder, 1994; Powell and Grimmer, 2016; Fouirnaies and Hall, 2018). A less studied but no less important source is the informal power derived from social connections. 

Social connections can shape the extent to which a legislator is sympathetic towards an interest group by, for instance, facilitating transmission and acquisition of relevant information between legislators. Yet, the bulk of the formal literature on interest groups conceives legislators as asocial agents and ignores connections of friendship and patronage between them. Battaglini and Patacchini (2018) is the first and only paper in the literature on interest groups that formally models legislators as social agents, and provides empirical evidence that campaign contributions by interest groups to legislators depend on the pattern of inter-personal influence between them. One would intuitively expect polarization between parties may impact inter-personal influence among legislators. Hence, it is natural to ask: does polarization between political parties benefit or hurt an interest group when we conceive of legislators as social agents? We investigate this question by suitably modifying the model in Battaglini and Patacchini (2018) to account for polarization between parties.

Our baseline model involves two parties, one ideologically aligned with the interest group and the other ideologically against the interest group. We simplify the decision of legislators by assuming they have to vote for one of two policies. The interest group has a strict preference for one of the policies but does not know the preferences of legislators with certainty.\footnote{One source of this uncertainty can be the unobserved personal convictions of the legislator (Ansolabehere et al., 2001). We further assume this uncertainty is large enough such that the interest group cannot guarantee a legislator will surely vote in its favor even if the interest group invests all of its budget towards the legislator.} It credibly promises monetary investments towards legislators conditional on voting for its preferred policy. Following Battaglini and Patacchini (2018), we assume the interest group allocates its limited budget to maximize the expected vote share for its preferred policy. Legislators derive resource-utility based on payments from the interest group but suffer an ideological-disutility upon voting against their party's ideology. The magnitude of this disutility serves as a measure of ideological polarization between the parties. 
	
The key element of the model is the interdependence in voting decisions of legislators. Specifically, we assume a legislator can be susceptible to another legislator belonging to either party in the sense that they may value voting in line with the latter (e.g., to seek patronage or to maintain social ties). The overall pattern of who is susceptible to whom is formally represented as a network with \textit{directed} links that allows for unilateral or mutual susceptibility between a pair of legislators. The network structure determines the \textit{influence} of each legislator, which can be interpreted as an indicator of ``key" legislators.\footnote{In our model, influence boils down to Bonacich centrality if all links are undirected, i.e., susceptibility is always mutual.} We refer to the aggregate influence of all legislators belonging to a party as the influence of the party. The influence of each legislator, and thus the influence of each party, is determined solely by the network structure and is independent of the level of ideological polarization between the parties.
	
The relative equilibrium investment towards any pair of legislators depends on their relative influence. As the influence of legislators is determined solely by the network structure, equilibrium investments are independent of ideological polarization between the parties. The equilibrium probability of a legislator voting in favor of the interest group is increasing in the investment by the interest group, and increasing (decreasing) in the level of ideological polarization if the legislator belongs to the party in favor of (against) the interest group. Consequently, the equilibrium expected vote share in favor of the interest group comprises two components. The first component depends on investments made by the interest group and is independent of ideological polarization. The second component depends on ideological polarization between the parties and the relative influence of the parties. The size and sign of the marginal impact of ideological polarization thus depend on the relative influence of the parties. 
	
The key result from the baseline model is that increase in ideological polarization between the parties benefits the interest group if and only if the party in favor of the interest group is relatively more influential. While the two major political parties in the United States currently seem to be as ideologically polarized as ever, political scientists view increasing affective polarization -- ill will, animus, or antipathy towards members of the opposite party -- as one of the most striking developments in U.S. politics (Mason, 2015; Rogowski and Sutherland, 2016; Iyengar and Krupenkin, 2018; Iyengar et al., 2019; Boxell et al., 2022). We next explore the impact of introducing affective polarization in the baseline model.\footnote{The bulk of existing literature has focused on affective polarization in the general public. To the extent that political elites exist in the same social context as the general public, and are drawn from the same population, it is conceivable that they are not completely immune to the broader social, economic, and political forces that seem to have fueled affective polarization among the masses. We are aware of only one published study that finds evidence of affective polarization among political elites. Using survey data from delegates to presidential nominating conventions, Enders (2021) finds that even the political elites ``exhibit extreme emotional reactions toward political groups just like their mass counterparts".} 
	
Affective polarization can be conceptualized \textit{as if} legislators value distinguishing their voting decisions from legislators in the ideologically opposite party (Druckman et al., 2021). As such, affective polarization effectively creates negative cross-party linkages and consequently modifies the influence of legislators. We formalize affective polarization by assuming a legislator suffers an affective-disutility when their voting decision coincides with a legislator in the opposite party. The magnitude of this affective-disutility serves as the measure of affective polarization between the parties. 
	
The relative equilibrium investment towards any pair of legislators once again depends on their relative influence, but the influence of a legislator depends on both the network structure and the level of affective polarization. The marginal impact of affective polarization on equilibrium expected vote share in favor of the interest group is therefore more nuanced than that of ideological polarization. In modifying the influence of legislators, affective polarization alters the effectiveness of investments by the interest group in incentivizing legislators to vote for its preferred policy. Further, by modifying the influence of each legislator, affective polarization also modifies the relative influence of parties. 
	
We show that sufficiently small levels of affective polarization reduce the effectiveness of investments by the interest group, and thus unambiguously hurt the interest group (relative to the absence of affective polarization). In contrast, sufficiently high levels of affective polarization can benefit the interest group depending on the network structure and the level of ideological polarization. This is possible, for instance, when ideological polarization is sufficiently high and the party in favor of the interest group is relatively more influential in the absence of affective polarization. In summary, our results suggest high levels of ideological or affective polarization are more likely to benefit the interest group when the party in favor of the interest group is relatively more influential.\\
	
%%%%%%%%%%%%%
%% Related Literature %%
%%%%%%%%%%%%%
	
\noindent {\bf Related literature.} Our work contributes to the literature on influence activities by interest groups (Snyder, 1991; Austen-Smith and Wright, 1992; Grossman and Helpman, 1994; Ansolabehere et al., 2003; Stratmann, 2002;  Baumgartner et al., 2009; Dekel et al., 2009; Lessig, 2011; Vidal et al., 2012; Bertrand et al., 2014; Schnakenberg, 2017; McKay, 2018). It also contributes to the literature on how networks among legislators impact legislative decisions (Rice, 1927; Routt, 1938; Truman, 1951; Arnold et al., 2000; Fowler, 2006; Cohen and Malloy, 2014; Harmon et al., 2019). In particular, our work complements Battaglini and Patacchini (2018), the first paper that provides a formal model to bridge these two historically distinct literatures.

Battaglini and Patacchini (2018), henceforth BP2018, examine competition between multiple interest groups in influencing the voting decisions of socially connected legislators. BP2018 theoretically shows that investments are proportional to Bonacich centralities of legislators under a wide variety of model specifications, and provides supporting empirical evidence. BP2018 shows that two competing interest groups with symmetric budgets nullify each other's impact in so far as legislators' voting decisions are concerned. This is not the case with asymmetric budgets. To the extent that some asymmetry between interest groups is a plausible assumption, our modeling choice to focus on one interest group can be viewed as an extreme form of asymmetry that helps obtain sharp comparative statics results regarding the impact of polarization.

Our paper focuses on identifying the conditions that benefit the interest group. Towards this end, we adapt the model in BP2018 to make salient the notions of  ``party'', ``ideology'', and ``affect''.\footnote{BP2018, however, controls for party affiliation of legislators in the empirical analysis.} These notions are central to our analysis as reflected in all the comparative statics results. In particular, we show that an \textit{aggregate} level feature of the network -- relative influence of parties -- is the key determinant of whether and when the interest group can exploit increasing polarization between two parties to its advantage. Further, our work offers a way to formalize affective polarization and clarifies how its impact on the interest group differs from that of ideological polarization.
	
Our work broadly relates also to the literature on informational lobbying where the tool of influence is information rather than money. The motivation behind several papers on informational lobbying is to explain why lobbies often target some specific legislators and provide them information (Caillaud and Tirole, 2007; Schnakenberg, 2017; Awad, 2020). Schnakenberg (2017) provides an illuminating discussion that casts these studies as attempts to resolve a puzzle. The puzzle arises from Hall and Deardorff (2006) that proposes a lobbyist primarily provides a legislative subsidy to allied legislators -- those who need no persuasion -- in the form of labor, resources, and time needed to develop customized issue-specific information and arguments to advance their mutually shared goals. Schnakenberg (2017) questions why do lobbyists then spend the bulk of their time and resources in constructing persuasive arguments and provide it to allied legislators who need no persuasion. While the various models of informational lobbying vary in their details, the common insight from these models is that information provided by a lobbyist assists the targeted legislators in persuading less sympathetic legislators. 
	
Our paper differs from such studies in its approach and objectives. We assume legislators care about their actions, whereas informational lobbying models assume they care about the policy outcome. Legislators are effectively conceived as asocial agents in informational lobbying models. We try to account for the context in which they operate via the patterns of inter-personal susceptibility. Belief-updating about the ``state of the world" by all legislators in response to the information provided by the lobbyist to the targeted legislators determines whether the lobby achieves its objective. We instead provide a preference-based account where the interest group exploits the inter-personal influence among legislators in order to adequately incentivize them to vote for its preferred policy. It is, perhaps, the search for belief-based explanations that lead informational lobbying models to conceive legislators as asocial agents. Finally, the role of affective polarization has not yet been explored in informational lobbying models.\\
	
%%%%%%%%%%%%%%
%% MODEL %%%%%%%
%%%%%%%%%%%%%%

\noindent \textbf{\Large 2. Baseline model}\\
	
\noindent Consider a set $L$ of $n \geq 2$ legislators. Each legislator belongs to a party $P \in \{F, A\}$. Let $n_{F} \ge 1$ and $n_{A} \ge 1$ denote the number of legislators in party $F$ and party $A$, respectively. Each legislator has to vote for a policy $p \in \{f, a\}$. There exists an interest group that strictly favors policy $f$ over policy $a$, and credibly promises investments towards legislators if they vote for its favored policy, i.e., policy $f$. Party $F$ is ideologically inclined towards policy $f$, whereas party $A$ is ideologically inclined towards policy $a$.
	
We seek to capture the idea that a legislator may value voting on the same policy as another legislator, but the converse need not necessarily be true. We, therefore, assume links are directed, and allow for both within-party and cross-party links. A directed link from legislator $i$ to legislator $j$ indicates $i$ values voting for the policy that $j$ votes for, whether it is policy $f$ or policy $a$. Thus, the presence of a directed link $\{\vec{ij}\}$ -- from $i$ to $j$ -- indicates two equivalent notions in the context of our model: whether $i$ is directly \textit{susceptible} to $j$, and whether $j$ has direct \textit{influence} over $i$.  
	
Formally, the overall pattern of who is susceptible to whom is represented by the $n\times n$ matrix $\mathbb{G} = [g_{ij}]_{n \times n}$, where $g_{ij}$ is 1 if legislator $i$ values voting like legislator $j$, and 0 otherwise.\footnote{Legislator $i$ can be indirectly susceptible to legislator $j$, even when not directly susceptible, if there exists a walk in the network from $i$ to $j$. Note that our analysis is valid for $g_{ij} \in [0, 1]$ where $g_{ij}$ can be interpreted as how strongly $i$ values voting like legislator $j$. We assume throughout that $g_{ii} = 0$ for every legislator, i.e., there are no self-loops in the network.} Since links are directed, it is possible that legislator $i$ values voting like legislator $j$, but legislator $j$ does not value voting like legislator $i$ (i.e., $g_{ij}$ need not be equal to $g_{ji}$). 
	
The overall network $\mathbb{G}$ is a $n \times n$ square matrix that can be viewed as a $2 \times 2$ \textit{block} matrix 
\begin{align}
\begin{aligned}
\quad\mathbb{G} = \left[ 
\begin{array}{c|c} 
\mathbb{G}_{FF} & \mathbb{G}_{FA} \\ 
\hline
\mathbb{G}_{AF} & \mathbb{G}_{AA} \\
\end{array} \right]
\end{aligned}
\end{align}
where the $n_{F} \times n_{F}$ matrix $\mathbb{G}_{FF}$ represents the pattern of directed links among legislators in party $F$. Similarly, the  $n_A \times n_A$ matrix $\mathbb{G}_{AA}$ represents the pattern of directed links among legislators in party $A$. The $n_F \times n_A$ matrix $\mathbb{G}_{FA}$ represents the pattern of directed links from legislators in party F to legislators in party A, and similarly, $n_A \times n_F$ matrix $\mathbb{G}_{AF}$ represents the pattern of directed links from legislators in party A to legislators in party F. As links are directed, the matrix $\mathbb{G}$ may differ from its transpose, $\mathbb{G}^{\top}$.\\
	
%%%%%%%%%%%%%%%%
%%%%%%%%%%%%%%%%
	
\noindent \textbf{\large 2.1. Objective of legislators}\\
	
\noindent In specifying the objective of legislators we assume they care about their actions -- i.e., which policy they vote for -- rather than the final outcome.\footnote{This assumption may be particularly suitable for large legislative bodies when each legislator anticipates their likelihood of being pivotal is very low.} The overall utility of a legislator comprises of three main components that relate to their ideological inclination, payment from the interest group, and the network effect. Specifically, the total utility of legislator $i$ belonging to party $F$ upon voting for policy $p \in \{f,a\}$ is
\begin{align}
U_{i}(p; F) = 
\begin{dcases}
~0 ~+~ u(m_{i}) ~+~ \delta~\sum\limits_{j \in L} g_{ij}~v_{j}(f) ~+~ \epsilon_{if} & ~~~~ \text{if ~$p = f$}\\
-\sigma ~+~ 0 ~+~ \delta~\sum\limits_{j \in L} g_{ij}~v_{j}(a) ~+~ 0 & ~~~~ \text{if ~$p = a$}\\
\end{dcases}
\end{align}
	
As party $F$ is ideologically inclined towards policy $f$, a legislator in party $F$ suffers an ideological-disutility of magnitude $\sigma > 0$ upon voting for policy $a$ but not upon voting for policy $f$. Since the interest group promises payments conditional on voting for policy $f$, a legislator in party $F$ obtains no resource-utility upon voting for policy $a$. In contrast, if a legislator votes for policy $f$, they obtain resource-utility $u(m_i)$, where $m_i$ denotes the payment to legislator $i$ promised by the interest group conditional on voting for policy $f$.\footnote{Our analysis is also valid for heterogeneous resource utility functions of legislators as long as the utility function $u_{i}(m_{i})$ of any legislator $i$ is continuous, strictly increasing, twice continuously differentiable, and strictly concave in $m_{i}$.} We assume the resource-utility function is continuous, strictly increasing, twice continuously differentiable, and strictly concave in $m_{i}$, with $u_{i}' \rightarrow \infty$ as $m_{i} \rightarrow 0$. 
	
The next component of the overall utility denotes the network-utility. The indicator variable $v_{j}(p)$ takes the value 1 if legislator $j$ votes for policy $p$, and zero otherwise. The parameter $\delta > 0$ denotes the network-utility derived by legislator $i$ per legislator that $i$ is directly susceptible to and who votes for the same policy as legislator $i$. We assume $\delta$ does not depend on the policy. This helps capture the idea that network-utility arises from voting for the same policy, which may be either policy $f$ or policy $a$.
	
The final component indicates a private exogenous preference shock that is realized after the interest group announces the vector of investments toward legislators but before the legislators vote. The preference shock captures all additional relevant information to assess the utility a legislator will derive if they vote for a policy. The interest group is thus uncertain about the precise preferences of legislators. The source of this uncertainty could be lack of knowledge of other factors that impact the decision of a legislator such as preferences of their constituents or unobservable personal convictions of the legislator. Without loss of generality, we assume the preference shocks provide information relevant only to policy $f$. The preference shocks are independently and identically distributed across legislators, and their distribution is assumed to be common knowledge. Specifically, they are uniformly distributed over the interval $[-\frac{1}{2\theta}, \frac{1}{2\theta}]$ with density $\theta > 0$ and mean zero.
	
The overall utility specification of legislators in party $A$ is analogous to that of legislators in party $F$. The only difference is that a legislator in party $A$ suffers an ideological-disutility upon voting for policy $f$, but not upon voting for policy $a$. The parameter $\sigma$ may thus be interpreted as indicating the extent of \textit{ideological polarization} between the two parties. The components corresponding to resource-utility, network-utility, and preference shocks are specified in the same way as for legislators in party $F$. Specifically, the utility of legislator $i$ belonging to party $A$ upon voting for policy $p \in \{f,a\}$ is 
\begin{align}
U_{i}(p; A) = 
\begin{dcases}
- \sigma ~+~ u(m_i) ~+~ \delta~\sum\limits_{j} g_{ij}~v_{j}(f) ~+~ \epsilon_{if} & ~~~~ \text{if ~$p = f$ } \\
~0 ~+~ 0 ~+~ \delta~\sum\limits_{j} g_{ij}~v_{j}(a) ~+~ 0  & ~~~~  \text{if ~$p = a$ }\\
\end{dcases}
\end{align}
	
\bigskip
	
%%%%%%%%%%%%%%%%%
%%%%%%%%%%%%%%%%%
	
\noindent \textbf{\large 2.2. Objective of the interest group}\\
	
\noindent The \textit{ex-ante} probability that legislator $i$ votes in favor of the interest group -- after the investments are announced by the interest group but before the private preference shocks are observed -- is the expected value of $v_{i}(f)$ given the distribution of preference shocks. We henceforth denote this expectation ${E}\big(v_{i}(f)\big)$ by $q_i$. Following BP2018, we restrict attention to pure investment strategies by the interest group and assume it allocates its resource of size $M > 0$ among the legislators to maximize the expected vote share for its favored policy. 
	
Let ${\bf M}$ denote the set of all feasible investment vectors ${\bf m} = (m_{1}, m_{2}, \ldots, m_{n})$ such that each $m_i \ge 0$ and $\sum\limits_{i = 1}^{n} m_i \le M$. Formally, the interest group announces a feasible payment vector ${\bf m}\in {\bf M}$ that maximizes
\begin{align} 
\mathcal{Q}(\mathbf{m}) ~=~ \sum\limits_{i =1}^{n} q_{i}(\mathbf{m}).
\end{align}
	
Arguably, if the outcome of voting is determined by majority rule, then the interest group may seek to maximize the expected probability of a majority of legislators voting for its preferred policy. We nevertheless use expected vote share as the objective of the interest group because \textit{comparative statics} results are extremely difficult to obtain under qualified majority voting rules.\footnote{Using the arguments in BP2018 (Section V. D), \textit{existence} of an equilibrium can be established.} In general, expected vote share is the technically justifiable objective function for the interest group if the outcome is determined by \textit{Random Dictatorship} wherein (a) each legislator would be equally likely to be selected as the dictator, and (b) the dictator's vote would determine the outcome. Further, if the outcome is determined by majority voting, then the expected vote share can serve as a plausible objective function for the interest group if the underlying parameters of the model are such that voting probabilities of every legislator are in a neighborhood of $\frac{1}{2}$.\footnote{See the Appendix for further discussion.} Hence, our results are most informative in contexts where every legislator in "on the fence".

It is worth noting that our model departs from BP2018 in three main ways. First, we allow for directed links between legislators whereas BP2018 assumes links are undirected. Second, we explicitly account for polarization between the two parties. Third, we restrict attention to one interest group whereas BP2018 allows for multiple interest groups. As mentioned before, BP2018 focuses on how the network structure shapes equilibrium investments by the interest groups towards legislators. In contrast, our focus lies in identifying the conditions under which the interest group can exploit polarization between the parties to its advantage.\\   
	
%%%%%%%%%%%%%%%%
%%  EQUILIBRIUM      %%%%
%%%%%%%%%%%%%%%%
	
\noindent {\bf \Large 3. Equilibrium and comparative statics}\\
	
\noindent An equilibrium $(\mathbf{m}^{\ast}, \mathbf{q}^{\ast})$ is a pair of an investment vector $\mathbf{m}^{\ast}$ announced by the interest group and a vector $\mathbf{q}^{\ast}$ of ex-ante probabilities of legislators voting in favor of the interest group where (i) $\mathbf{q}^{\ast}$ constitutes an equilibrium conditional on $\mathbf{m}^{\ast}$, and (ii) $\mathbf{m}^{\ast}$ maximizes the expected vote share in favor of the interest group. We first analyze the voting behavior of legislators conditional on an arbitrary vector of payments ${\bf m} = (m_{1}, \ldots, m_{i}, \ldots, m_{n})$ announced by the interest group. \textit{Conditional} on her private preference shock $\epsilon_{if}$, the expected utility of legislator $i \in F$ if she votes policy $p \in \{f, a\}$ will be 
\begin{align}
{E}\big(U_{i}(p; F|\epsilon_{if})\big) = 
\begin{dcases}
~0 ~+~ u(m_i) ~+~ \delta~\sum\limits_{j} g_{ij}~q_{j}  ~+~ \epsilon_{if} & ~~~~ \text{if ~$p = f$}\\
-\sigma ~+~ \delta~\sum\limits_{j} g_{ij}~(1-q_{j})  & ~~~~ \text{if ~$p = a$}
\end{dcases}
\end{align}
	
As described earlier, the expected value of $v_{j}(f)$ from the perspective of legislator $i \ne j$ is $q_{j}$ -- the probability that legislator $j$ votes in favor of the interest group -- since preference shocks are observed privately. Consequently, the probability that legislator $j$ votes against the interest group will be $(1-q_{j})$. Further, the expected value of the preference shock conditional on observing the preference shock will simply be the observed preference shock. Hence, \textit{conditional} on her preference shock, the expected utility of legislator $i \in A$ if she votes for policy $p$ will be
\begin{align}
{E}\big(U_{i}(p; A|\epsilon_{if})\big) = 
\begin{dcases}
-\sigma ~+~ u(m_{i}) ~+~ \delta~\sum\limits_{j} g_{ij}~q_{j}  ~+~ \epsilon_{if} & ~~~~ \text{if ~$p = f$ }\\
0 ~+~ \delta~\sum\limits_{j} g_{ij}~(1-q_{j}) & ~~~~ \text{if ~$p = a$ }\\
\end{dcases}
\end{align}
	
Conditional on the privately observed shock, a legislator votes for the interest group if her expected utility from voting for the interest group is at least as high as her expected utility from voting against the interest group, i.e., if 
\begin{align}
{E}\big(U_{i}(f ; P|\epsilon_{if})\big) ~\geq~ {E}\big(U_{i}(a ; P|\epsilon_{if})\big).
\end{align}

Consequently, the \textit{ex-ante probability} that a legislator votes for the interest group -- i.e., before she observes her private preference shock -- will be the probability that her private preference shock exceeds a certain threshold. The probability that legislator $i$ in party $P \in \{F, A\}$ votes in favor of the interest group turns out to be 
\begin{align}
q_{i}({\bf m}; P) ~&=~ \frac{1}{2} ~+~ \theta \Bigg(\Big(u(m_{i})~+~ \delta~ \sum\limits_{j \in L} g_{ij}~(2q_{j} - 1)\Big) ~+~ \sigma_P \Bigg),
\end{align}
\noindent where $\sigma_P = \sigma$ if $P = F$ and $\sigma_P = -\sigma$ if $P = A$.  
	
\noindent Following BP2018, we assume the underlying parameters of the model are such that voting probabilities lie in the \textit{interior} of the unit interval (henceforth, Assumption 1). This ensures the interest group is never sure about the voting choice of any legislator. Assumption 1 will hold, for instance, if the density of preference shocks, $\theta$, is small enough.\footnote{The Appendix provides the formal details behind Assumption 1.} 

Given the interdependence in voting probabilities of legislators, a vector of equilibrium voting probabilities of legislators conditional on an arbitrary vector of investments by the interest group will be a solution to the system of \textit{linear} equations generated by equation (8). The vector of voting probabilities in favor of the interest group is given by $~\mathbf{q} ~=~ (q_{1}, q_{2}, ~\cdots ~, q_{n})$. The system of simultaneous equations of the legislators' voting in favor of the interest group is as follows
\begin{align}
\mathbf{q}(\mathbf{m}) ~&=~ 
\begin{pmatrix} q_{1}(\mathbf{m}) \\ \vdots \\ q_{n}(\mathbf{m})\end{pmatrix} = 
\begin{pmatrix} \frac{1}{2} ~+~ \theta~ \big[ u(m_{1}) ~+~ \sigma +~\delta \sum\limits_{j} g_{1j}~(2q_{j}(\mathbf{m}) - 1) \big] \\ \vdots \\ \frac{1}{2} ~+~ \theta~\big[ u(m_{n}) ~-~ \sigma +~\delta \sum\limits_{j} g_{nj}~(2q_{j}(\mathbf{m}) - 1) \big] \end{pmatrix} \\ \notag
~&~\\
\mathbf{q}(\mathbf{m}) ~&=~  \frac{1}{2} \cdot \mathbf{1} ~+~ \theta\cdot \mathbf{u(m)} ~+~ \theta \cdot \bm{\sigma} ~+~ 2\theta \delta \cdot \mathbb{G}\cdot\mathbf{q}(\mathbf{m})  ~-~ \theta \delta \cdot\mathbb{G} \cdot \mathbf{1}
\end{align}

\noindent where $\mathbf{1}$ is the $n\times1$ vector of ones, $\mathbf{u}({\bf m}) = (u(m_{1}), u(m_{2}), ~\cdots ~ , u(m_{n}))$ is the vector of resource-utilities derived by legislators, $\bm{\sigma} ~=~ (\sigma , \sigma , ~\cdots ~, - \sigma , - \sigma)$ is the vector that accounts for ideological inclinations of legislators (with the positive entries corresponding to legislators in the party $F$), and $\mathbb{G}$ is the network among all legislators. 

We establish the existence of a unique vector of voting probabilities that solves the above system of linear equations following the proof of Proposition 1 in BP2018. Note that this system of linear equations induces a continuous mapping of probability vectors from $[0,1]^{n}$ to $(0,1)^{n}$. The domain of this mapping is a compact and convex set, the co-domain is a subset of the domain, and the mapping is continuous as it is linear. Hence, Brouwer's fixed-point theorem ensures a solution exists. The following assumption ensures the solution is unique.\\

\noindent {\bf [Assumption 2]} ~For any feasible network $\mathbb{G}$, the matrix $(\mathds{I} - 2\delta \theta \mathbb{G})$ is invertible.

\bigskip

\noindent Invertibility is ensured if $\beta = 2\delta \theta$ is sufficiently small. A sufficient condition for invertibility is that the largest eigenvalue of $\mathbb{G}$, $\bar{\lambda}(\mathbb{G})$, is such that $\beta \bar{\lambda}(\mathbb{G}) < 1$ (Ballester et al., 2006; BP2018). We assume $\beta n < 1$ for the remainder of this section, which suffices to ensure invertibility of every feasible network.
When both assumptions 1 and 2 hold, equation (10) can be rearranged to obtain
\begin{align}
(\mathds{I} - \beta \mathbb{G}) \cdot \mathbf{q(m)} ~&=~ \frac{1}{2} \cdot (\mathds{I} - \beta \mathbb{G})\cdot\mathbf{1} ~+~ \theta \cdot \big( \mathbf{u(m)} ~+~ \bm{\sigma} \big)
\end{align}
\noindent where $\mathds{I}$ is the $n \times n$ identity matrix, $\beta = 2\delta \theta \in (0,1)$. The unique equilibrium vector of voting probabilities conditional on an arbitrary investment vector will thus be 
\begin{align}
\mathbf{q}^{\ast}({\bf m}) ~=~ \frac{1}{2} \cdot \mathbf{1} ~+~ \theta \cdot (\mathds{I} - \beta \mathbb{G})^{-1} \cdot \big( \mathbf{u}({\bf m})  ~+~ \bm{\sigma} \big).
\end{align}
	
The above equation implies that a legislator's probability of voting in favor of the interest group is increasing in the investment by the interest group, and increasing (decreasing) in the level of ideological polarization if the legislator belongs to the party in favor of (against) the interest group. It also implies the expected vote share in favor of the interest group as a function of an arbitrary investment vector ${\bf m}$ is 
\begin{align}
\mathcal{Q}^{\ast}(\mathbf{m}) ~=~ \sum_{i \in L} q^{\ast}_{i}({\bf m}) ~&=~ \big(\mathbf{q}^{\ast}({\bf m})\big)^{\top} \cdot \mathbf{1},
\end{align}
	
\noindent where $\big(\mathbf{q}^*({\bf m})\big)^{\top}$ is the transpose of the vector of voting probabilities and $\mathbf{1}$ is the $n \times 1$ vector of ones. We are now ready to state the equilibrium existence and uniqueness, a result analogous to Proposition 1 in BP2018.\\
	
%%%%%%%%%%%%%%%
%%  PROPOSITION 1      %%
%%%%%%%%%%%%%%%
	
\noindent {\bf Proposition 1.} \textit{There exists a unique equilibrium, $\big(\mathbf{m}^{\ast}, \mathbf{q}^{\ast}(\mathbf{m}^{\ast})\big)$.}\\
	
\noindent {\bf Proof.} The expected vote share $\mathcal{Q}^*(\mathbf{m})$ is continuous and strictly concave in the investment vector ${\bf m}$ due to the continuity and strict concavity of the resource-utility functions of legislators. Weierstrass theorem ensures the existence of a maximizer due to the compactness of the set of feasible investment vectors and continuity of $\mathcal{Q}^*({\bf m})$. Strict concavity of $\mathcal{Q}^*({\bf m})$ in the investment vector ${\bf m}$ ensures uniqueness. $~\blacksquare$\\
	
%%%%%%%%%%%%%%%%
%%%%%%%%%%%%%%%%
	
\noindent {\bf \large 3.1. Ideological Polarization}\\
	
\noindent Using equation (13), the expected vote share in favor of the interest group as a function of an arbitrary investment vector ${\bf m}$ can be expressed as
\begin{align}
\mathcal{Q}^{\ast}(\mathbf{m}) ~=~ \big(\mathbf{q}^{\ast}({\bf m})\big)^{\top} \cdot \mathbf{1}~=~ \frac{n}{2} ~+~ \theta \big(\mathbf{u}({\bf m})  ~+~ \bm{\sigma}\big)^{\top} \cdot \bm{\mathcal{I}},
\end{align}
	
\noindent where the vector $\bm{\mathcal{I}} = (\mathit{I}_{1}, \mathit{I}_{2},\cdots, \mathit{I}_{n})^{\top} = (\mathds{I} - \beta \mathbb{G}^{\top})^{-1} \cdot \mathbf{1}$. The entries in the vector $\bm{\mathcal{I}}$ can be interpreted as the overall \textit{influence} of each legislator given the pattern of linkages in the network.\footnote{Assumption A2 ensures the matrix $(\mathds{I} - \beta \mathbb{G}^{\top})$ is invertible.} Intuitively, the influence of any legislator $i$ depends on the influence of all other legislators that are susceptible to legislator $i$. Specifically,
\begin{align*}
\mathit{I}_{i} ~=~ 1 + \beta \cdot \sum\limits_{j = 1}^{n} g_{ji} \cdot \mathit{I}_{j}
\end{align*}

\noindent For a given network $\mathbb{G}$ and decay factor $\beta$, the influence vector is thus given by
\begin{align*}
\bm{\mathcal{I}}(\beta, \mathbb{G})  ~=~ (\mathds{I} - \beta \mathbb{G}^{\top})^{-1} \cdot \mathbf{1} ~=~ \sum\limits_{k = 0}^{+\infty } \beta^{k} \cdot (\mathbb{G}^{\top})^{k}\cdot \mathbf{1},
\end{align*}
\noindent In the context of our model, the decay factor $\beta$ is $2 \delta \theta$. In the $n \times n$ matrix $(\mathds{I} - \beta \mathbb{G}^{\top})^{-1} = \big[ x_{ij} \big]$, the element $x_{ij} \ge 0$ denotes the total number of discounted walks from $j$ to $i$, where the discounting factors are the powers of the decay factor $\beta$.\footnote{When $\mathbb{G} = [0]_{n \times n}$, $\bm{\mathcal{I}} = (1, \ldots, 1, \ldots 1)$. Hence, $x_{ii} \ge 1$ for every $i \in L$ under any feasible network.} The entry $x_{ij}$ can be interpreted as the overall influence of legislator $i$ on legislator $j$ that accounts for both the direct as well as the indirect influence of $i$ on $j$. The overall \textit{influence} of any legislator $i$ on \textit{all} legislators is thus the \textit{row sum} $\mathit{I}_{i} = \sum_{j=1}^{n} ~x_{ij} \ge 1$. It is worth emphasizing that the influence of a legislator is solely determined by the structure of the network. If links are undirected, then the overall influence of a legislator equals the Bonacich centrality of the legislator, as in BP2018.

The expected vote share in favor of the interest group as a function of an arbitrary investment vector ${\bf m}$ can be rewritten as
\begin{align}
\mathcal{Q}^{\ast}(\mathbf{m}) ~=~ \frac{n}{2} ~+~ \theta \big(\sum\limits_{i \in L} \mathit{I}_{i} \cdot u(m_{i}) \big) ~+~ \theta \sigma \cdot \big(\mathit{I}_{F} ~-~ \mathit{I}_{A}\big),
\end{align}
	
\noindent where $I_i$ is the influence of legislator $i \in L$, and $\mathit{I}_{P} = \sum_{i \in P} I_{i}$ is the sum of influences of all legislators in party $P \in \{F, A\}$. Thus, the expected vote share in favor of the interest group comprises of three main components. The first term is the expected vote share in the absence of investments and ideological polarization. The second component depends on investments by the interest group but is independent of ideological polarization. The final component depends on ideological polarization between the parties and the relative influence of the parties. We, therefore, obtain the following result.\\
	
\noindent {\bf Corollary 1.} \textit{The interest group chooses the investment vector that maximizes influence-weighted sum of legislators' resource-utilities. Consequently, equilibrium investments by the interest group are independent of the extent of ideological polarization between parties.}\\
	
\noindent The result follows from the feature that the influence of legislators is determined solely by the network structure. It is independent of ideological polarization $\sigma$ between the parties. The optimal investment vector is such that the influence-weighted marginal resource-utilities are identical for all legislators. Formally, $I_{i} \cdot u'(m_{i}^{\ast}) = I_{j} \cdot u'(m_{j}^{\ast})$ for any two legislators $i$ and $j$. 
	
While ideological polarization does not affect investments toward legislators, it does affect the equilibrium expected vote share in favor of the interest group. Hence, increase in ideological polarization can impact the interest group. The following proposition highlights that the relative influence of the two parties determines the marginal impact of ideological polarization on the interest group.\\
	
%%%%%%%%%%%%%%%
%%  PROPOSITION 2     %%
%%%%%%%%%%%%%%%
	
\noindent {\bf Proposition 2.} \textit{An increase in ideological polarization benefits the interest group if and only if the party in favor of the interest group is relatively more influential}.\\
	
\noindent {\bf Proof.} Using (15), the equilibrium value of the interest group's objective is
\begin{align}
\mathcal{Q}^{\ast}(\mathbf{m}^{\ast}) ~=~ \frac{n}{2} ~+~ \theta \big(\sum\limits_{i \in F, A} 
\mathit{I}_{i} \cdot u(m_{i}^{\ast}) \big) ~+~ \theta \sigma \cdot \big(\mathit{I}_{F} ~-~ \mathit{I}_{A}\big)
\end{align}
	
\noindent The result follows directly from the observation that $\frac{\partial \mathcal{Q}^{\ast}}{\partial \sigma} \geq 0$ if and only if $\mathit{I}_{F} \ge \mathit{I}_{A}$. $~\blacksquare$
	
\bigskip
	
This result highlights the crucial role of relative influence of the two parties in determining whether greater ideological polarization between parties benefits the interest group. To further see the importance of this proposition, it is helpful to contrast it with a model that ignores the network among legislators. Such a model is a special case of our model, and corresponds to $\mathbb{G}$ being the empty network. The influence of every legislator will then be unity, and the influence of a party will simply be the number of legislators in the party. Consequently, if party $A$ has relatively fewer legislators, then an increase in ideological polarization can never hurt the interest group. In this sense, the novelty of Proposition 2 lies in demonstrating that greater influence of the party against the interest group can compensate for its potential numerical disadvantage in limiting the vote share in favor of the interest group.\\
	
%%%%%%%%%%%%
%%%%%%%%%%%%
	
\noindent \textbf{\large 3.2. Impact of changes in network structure}\\
	
\noindent We shall say a change in the network from $\mathbb{G}$ to $\mathbb{G}'$ benefits the interest group if the equilibrium expected vote share under $\mathbb{G}'$ is relatively higher, i.e., if $\mathcal{Q}^{\ast}(\mathbb{G}') \ge \mathcal{Q}^{\ast}(\mathbb{G})$.\footnote{We suppress the dependence on other primitives for notational convenience.} We say network $\mathbb{G}^{+} = [g^{+}_{ij}]$ is \textit{stronger} than network $\mathbb{G} = [g_{ij}]$ if every $g^{+}_{ij} \ge g_{ij}$, and at least one inequality is strict. When the network becomes stronger, at least one legislator becomes directly susceptible to at least one other legislator to whom they were not previously susceptible.\\
	
\noindent {\bf Lemma 1.} \textit{If the strength of the network increases, then the influence of every legislator (weakly or strictly) increases.}\\
	
\noindent {\bf Proof outline.} Consider a change in the network whereby legislator $i$ becomes directly susceptible to legislator $j \ne i$. We show in the Appendix that the change in influence of any legislator $k \in L$ is 
\begin{align}
\Delta \mathit{I}_{k} ~\equiv~ \mathit{I}_{k}^{+}  ~-~ \mathit{I}_{k} ~=~ \big( \frac{\beta \cdot \mathit{I}_{i}}{1 - \beta \cdot x_{ij}} \big) \cdot x_{kj}
\end{align}
\noindent which is increasing in the prior influence of legislator $i$, the prior influence of legislator $i$ on legislator $j$, and the prior influence of legislator $k$ on legislator $j$. As described in Section 3.1, $\beta > 0$, and under any network, every $x_{ij} \ge 0$ and $I_{i} \ge 1$ for every $i \in L$. We show in the Appendix that the term $(1 - \beta x_{ij})$ is strictly positive as well. In general, the addition of any new link may alter the relative influence of various legislators but does not decrease the absolute influence of any legislator. $~\blacksquare$\\
	
Consider the change in network from $\mathbb{G}$ to a relatively stronger network $\mathbb{G}^{+}$.  The \textit{change in the influence} of party $P \in \{F, A\}$ will be $\Delta \mathit{I}_{P} ~\equiv~ \mathit{I}_{P}^{+} ~-~ \mathit{I}_{P} ~=~ \sum\limits_{i \in P} ~(I_{i}^{+} ~-~ I_{i})$. This change in network benefits the interest group if 
\begin{align}
\mathcal{Q}^{\ast}(\mathbb{G}^+) ~-~ \mathcal{Q}^{\ast}(\mathbb{G}) ~=~ \sum\limits_{i \in L} ~\big(I_{i}^{+} \cdot u(m_{i}^{+ \ast}) ~-~ I_{i} \cdot u(m_{i}^{\ast}) \big) ~+~ \sigma \cdot (\Delta I_F ~-~ \Delta I_A) ~\ge 0,	
\end{align}
	
\noindent where $\mathbf{m}^{\ast} = (m_1^{\ast}, \ldots, m_i^{\ast}, \ldots, m_n^{\ast})$ and $\mathbf{m}^{+\ast} = (m_1^{+*}, \ldots, m_i^{+{\ast}}, \ldots, m_n^{+{\ast}})$ denote the equilibrium investment vectors under $\mathbb{G}$ and $\mathbb{G}^{+}$, respectively. The first part of the above inequality indicates the investment effect, and the second part is the ideological polarization effect.\\
	
%%%%%%%%%%%%%%%
%%  PROPOSITION 3      %%
%%%%%%%%%%%%%%%
	
\noindent {\bf Proposition 3.} \textit{Consider a change in the network such that the resulting change in the influence of party $P \in \{F, A\}$ is $\Delta I_{P}$.}\\
\noindent {\bf (a)} \textit{If $\Delta I_{F} \ge \Delta I_{A} \geq 0$, then the change in network benefits the interest group.}\\
\noindent {\bf (b)} \textit{If $\Delta I_{A} > \Delta I_{F} \geq 0$, then the change in network benefits the interest group only if ideological polarization $\sigma$ is below a threshold $\hat{\sigma}$, where}
\begin{align*}
\hat{\sigma} ~ = ~ \frac{\sum\limits_{i \in L} ~\big(I_{i}^{+} \cdot u(m_{i}^{+ \ast}) ~-~ I_{i} \cdot u(m_{i}^{\ast}) \big)}{\Delta I_{A} - \Delta I_{F}}.
\end{align*}
	
\noindent {\bf Proof outline.} The detailed proof is provided in the Appendix. In brief, the result follows from the investment effect being non-negative. The interest group can always tailor investments under the stronger network in such a way that, in equilibrium, the influence weighted sum of resource utilities of all legislators is relatively higher under the stronger network. The sign of the polarization effect depends on which party's influence increases more due to the change in network. If $\Delta I_F \ge \Delta I_A \ge 0$, then both investment and ideological polarization effects are non-negative. Consequently, a relatively greater change in the influence of party $F$ benefits the interest group at every level of ideological polarization. If $\Delta I_A > \Delta I_F \ge 0$, then the investment effect remains non-negative but the ideological polarization effect is strictly negative. Hence, a relatively greater increase in the influence of party $A$ benefits the interest group only if ideological polarization is sufficiently low. $~\blacksquare$\\
	
Intuitively, part (a) of Proposition 3 highlights the interest group can exploit the increase in the strength of the network to their advantage if the resulting change in the influence of the party in favor of the interest group is relatively greater. Part (b) highlights the interest group can exploit an increase in the strength of the network even if the resulting increase in the influence of party $A$ is relatively greater. However, this is only possible when ideological polarization between the parties is sufficiently low such that the ideological-disutility to legislators in party $A$ upon voting in favor of the interest group is not too large. It is only in such cases that the interest group can tailor higher investments towards at least some legislators in party $A$ that have sufficiently high influence to increase their probability, as well as of those legislators that are particularly susceptible to them, of voting in favor of the interest group.\\    
	
In the following, we present two examples. The first example illustrates all the above results. The second details a special case of the model with no cross-party links between legislators.\\
	
%%%%%%%%%%%%%%%
%%  EXAMPLE 1      %%
%%%%%%%%%%%%%%%
	
\noindent \textbf{Example 1.} \textit{Consider any network and suppose every legislator has the same increasing and concave resource-utility function $u(m_{i}) = \sqrt{m_{i}}$.} Let $I_i$ denote the influence of any legislator $i \in L$ which depends solely on the network structure. The influence of party $P \in \{F, A\}$ will be $I_{P} = \sum\limits_{j \in P} I_{j}$. Proposition 1 ensures existence of a unique equilibrium. Corollary 1 implies equilibrium investment on any legislator $i$ is independent of ideological polarization $\sigma$. Further, equilibrium investments will be such that the influence-weighted marginal resource-utilities will be equal across all legislators. Consequently, equilibrium investment toward any legislator $i \in L$ will be $m_{i}^{\ast} = \big(\frac{I_{i}^{2}}{\sum\limits_{j \in L}^{~^{~}} I_{j}^{2}} \big) \cdot M$. The equilibrium expected vote share in favor of the interest group will be
\begin{align*}
\mathcal{Q}^{\ast}(\mathbf{m}^{\ast}) ~=~ \frac{n}{2} ~+~ \theta \sum\limits_{i \in L} ~\big(\mathit{I}_{i} \cdot \sqrt{m_{i}^{\ast}} ~\big)  ~+~ \theta \sigma \cdot \big(\mathit{I}_{F} ~-~ \mathit{I}_{A}\big)
\end{align*}
	
\noindent In line with Proposition 2, an increase in ideological polarization $\sigma$ strictly benefits the interest group if and only if $\mathit{I}_{F} > \mathit{I}_{A}$.
	
Finally, we use a numerical example to demonstrate the impact of a change in the network on the interest group as highlighted in Proposition 3. Consider the network $\mathbb{G}$ in Figure 1(a) wherein $F2$ is the most influential legislator. The change in network from $\mathbb{G}$ to $\mathbb{G}^+$ involves $F2$ becoming directly susceptible to legislator $A2$, which in turn increases the influence of $A2$. Suppose $(\theta, \delta, M) =(0.03, 0.3, 100)$. In line with Lemma 1, this change in the network increases the influence of every legislator (Row 1 in Table 1). The investment towards $A2$ increases while the investment towards every other legislator decreases (Row 2). The change in network creates a relatively larger increase in the influence of party $A$  ($\Delta \mathit{I}_{A} = 0.0187 > \Delta \mathit{I}_{F} = 0$). Proposition 3(b) therefore suggests the change in network will strictly benefit the interest group if and only if ideological polarization $\sigma < \hat{\sigma} = 4.9442$. When $\sigma = 3$, the stronger network benefits the interest group as reflected in the expected vote share in favor of the interest group that increases by $0.0011$ (Row 3). In contrast, when $\sigma = 6$, the stronger network hurts the interest group with the expected vote share decreasing by $0.0005$ (Row 4 in Table 1).\\

%%%%%%%%%%%%%%%
%%  FIGURE 1      %%%%%%
%%%%%%%%%%%%%%%

\begin{figure}[]
\centering
\begin{subfigure}{.45\textwidth}
\centering
\includegraphics[width = 0.5\textwidth]{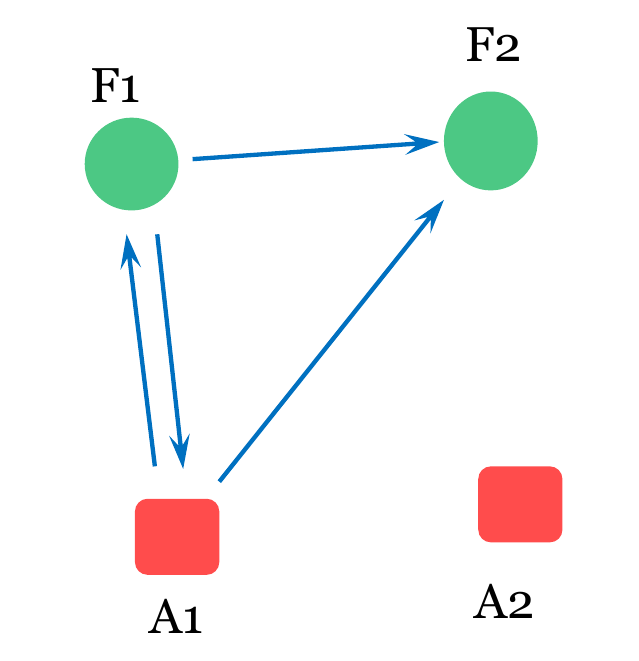}
\caption{$\mathbb{G}$}
\end{subfigure}
\begin{subfigure}{.45\textwidth}
\centering
\includegraphics[width=0.5\textwidth]{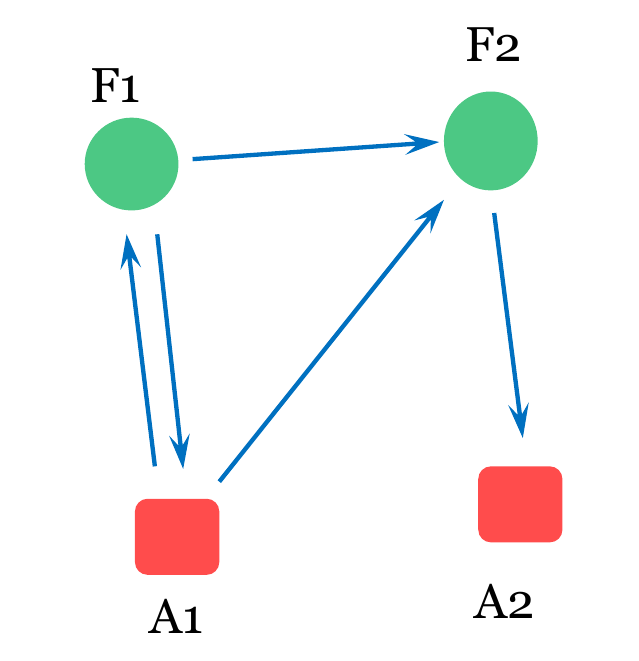}
\caption{$\mathbb{G}^{+}$}
\end{subfigure}
\vspace*{0.3cm}
\caption{Change in network from $\mathbb{G}$ to $\mathbb{G}^+$ (Example 1). The influence ranking of legislators is $F2 > F1=A1 > A2$ and $F2 > A2 > F1 = A1$ under $\mathbb{G}$ and $\mathbb{G}^+$, respectively.}
\vspace*{0.5cm}
\end{figure}

%%%%%%%%%%%%%%%
%% TABLE 1  %%
%%%%%%%%%%%%%%%

\begin{table} []
\caption{Impact of change in network from $\mathbb{G}$ to $\mathbb{G}^+$ (Example 1)}
\begin{center}
\begin{tabularx}{\textwidth}{c l  c c c c c c c c c }
\hline \hline
&{~~~~~~~~~}&&&&& \bf{~~~~F1~~~~} & \bf{~~~~F2~~~~}  & \bf{~~~~A1~~~~} & \bf{ ~~~~A2 ~~~~}&\bf{~~~~Total~~~~}  \\
\hline
&&&&&&&&&& \\
&$\Delta I$         &&&&& 0 & 0  & 0 & 0.0187 & 0.0187\\
&&&&&&&&&& \\
&$\Delta m^*$        &&&&& -0.2249 & -0.2331 & -0.2249 & 0.6829 & 0 \\
&&&&&&&&&& \\
&$\Delta q^*(\sigma=3)$ &&&&& -0.0007 & 0.0004 & -0.0007 & 0.0021 & 0.0011  \\
&&&&&&&&&&\\
&$\Delta q^*(\sigma=6)$    &&&&& -0.0007 &  -0.0012  & -0.0007 & 0.0021 & -0.0005  \\
&&&&&&&&&& \\
\hline
\hline
\end{tabularx}
\end{center}
\noindent {\footnotesize \textit{Notes.} The first row reports the change in influence and the second row reports the change in equilibrium investment. These changes are independent of ideological polarization $\sigma$. The third and fourth rows report the change in equilibrium probabilities of voting in favor of the interest group at $\sigma=3$ and $\sigma=6$, respectively.}\\
\end{table}
	
%%%%%%%%%%%%%%%
%%  EXAMPLE 2             %%
%%%%%%%%%%%%%%%
	
\noindent \textbf{Example 2.} \textit{Consider a special case with no cross-party links and undirected within-party links}. In the absence of cross-party links, no legislator can directly or indirectly influence any legislator in the opposite party. Due to links being undirected, the influence of a legislator will be the \textit{Bonacich centrality} of the legislator within one's own party network. Consequently, the influence of a party will be the sum of Bonacich centralities of all its legislators. 
	
The existence of a unique equilibrium is ensured by Proposition 1. Corollary 1 will imply the unique equilibrium investment vector (a) maximizes the Bonacich centrality weighted sum of resource utilities, and (b) is independent of the level of ideological polarization between the parties. Proposition 2 will imply an increase in ideological polarization will benefit the interest group if and only if the party in favor of the interest group is relatively more influential. Lemma 1 will imply an additional (undirected) link within a party increases the Bonacich centrality of every legislator in the party. Proposition 3 will imply (a) a stronger network in party $F$ will unambiguously benefit the interest group, whereas (b) a stronger network in party $A$ will benefit the interest group if and only if ideological polarization is sufficiently low. In this special case, using Corollary 1 and Lemma 1, it can also be shown that an increase in the strength of the network in a party increases total investment towards the party.\footnote{Lemma 1 implies a stronger network in a party $P$ will weakly or strictly increase Bonacich centrality of its legislators but not alter the Bonacich centrality of any legislator in party $P' \ne P$. Corollary 1 implies the equilibrium investment on every legislator in party $P'$ will be relatively lower under the relatively stronger network in party $P$. Hence, total investment towards party $P'$ will decrease. Consequently, given the fixed budget of the interest group, total investment towards party $P$ will increase.}\\
	
%%%%%%%%%%%%%%%%%%%%
%%  AFFECTIVE POLARIZATION  %%
%%%%%%%%%%%%%%%%%%%%
	
\noindent \textbf{\Large 4. Affective polarization}\\
	
\noindent We model affective polarization as an affective-disutility that a legislator suffers when a legislator in the opposite party votes for the same policy. This is equivalent to assuming each legislator in one party values distinguishing her vote from legislators in the other party. The size of the affective-disutility serves as a measure of affective polarization between the parties. For analytical convenience, we assume there are no cross-party links. Thus, the specification of overall utility of a legislator differs from the baseline model in two ways. First, legislators derive positive network-utility only from legislators within their own party. Second, a legislator suffers affective-disutility upon voting like a legislator in the other party. Specifically, we assume the total utility of legislator $i$ belonging to party $F$ upon voting for policy $p \in \{f, a\}$ is
\begin{align}
U_{i}(p; F) ~=~ 
\begin{dcases}
~~0 ~+~ u(m_{i}) ~+~ \delta \sum\limits_{j \in F} g_{ij}~v_{j}(f) ~-~ \alpha \sum\limits_{j \in A}~  v_{j}(f) ~+~ \epsilon_{if} & ~~~~ \text{if ~$ p= f$}\\
~~ -\sigma ~+~ 0 ~~~+~ \delta \sum\limits_{j \in F} g_{ij}~v_{j}(a) ~-~ \alpha \sum\limits_{j \in A}~v_{j}(a) & ~~~~\text{if ~$p = a$}
\end{dcases}
\end{align}

Analogously, the total utility of legislator $i$ belonging to party $A$ upon voting for policy $p \in \{f, a\}$ is
\begin{align}
U_{i}(p; A) ~=~ 
\begin{dcases}
-\sigma ~+~ u(m_{i}) ~+~ \delta \sum\limits_{j \in F} g_{ij}~v_{j}(f) ~-~ \alpha \sum\limits_{j \in A}~  v_{j}(f) ~+~ \epsilon_{if} & ~~~~ \text{if ~$ p= f$}\\
~	~0 ~+~~ 0 ~~~+~ \delta \sum\limits_{j \in A} g_{ij}~v_{j}(a) ~-~ \alpha \sum\limits_{j \in F}~v_{j}(a) & ~~~\text{if ~$p = a$}
\end{dcases}.
\end{align}

The parameter $\alpha>0$ captures the affective-disutility suffered by a legislator in party $P$ upon voting for policy $p$ per legislator in the other party who also votes for policy $p$. In other words, as opposed to the baseline model, the payoff of the legislator upon voting for policy $p$ is amended by a network disutility from those in the other party who also vote for policy $p$. The disutility term is $- \alpha \sum\limits_{j \in P}~  v_{j}(p)$.

Conditional on an investment vector ${\bf m}$, the probability that any legislator $i$ in party $P \in \{F,A\}$ votes in favor of the interest group is
\begin{align}
q_{i}({\bf m}; P) ~&=~ \frac{1}{2} ~+~ \theta \Big(u(m_{i})~+~ \delta~ \sum\limits_{j \in P} g_{ij}~(2q_{j} - 1) ~-~ \alpha \sum\limits_{j \notin P} (2q_{j} - 1) ~+~ \sigma_P \Big),
\end{align}
\noindent where $\sigma_P = \sigma$ if $P = F$ and $-\sigma$ if $P = A$. 
	
As in the baseline model, we restrict attention to environments where voting probabilities lie in the interior of the unit interval (Assumption 3). A vector of equilibrium voting probabilities of legislators conditional on an arbitrary vector of investments by the interest group is a solution to the system of \textit{linear} equations generated by equation (21). Brouwer's fixed-point theorem ensures a solution exists. In order to ensure uniqueness, we need an assumption analogous to Assumption 2 in the baseline model. Towards this end, it is helpful to first define the matrix
\begin{align}
\begin{aligned}
\quad \widehat{\mathbb{G}} ~=~ \left[ 
\begin{array}{c|c} 
\delta \cdot \mathbb{G}_{FF}& -\alpha \cdot \mathds{1}_{FA} \\ 
\hline
-\alpha \cdot \mathds{1}_{AF} & \delta \cdot \mathbb{G}_{AA} \\
\end{array} \right]
\end{aligned}
\end{align}
\noindent where the blocks on the main diagonal correspond to the networks within the two parties multiplied by the network-utility, and the off-diagonal blocks are matrices of ones multiplied by the affective-disutility having the appropriate order such that $\widehat{\mathbb{G}}$ is a symmetric $n \times n$ square matrix.\\
	
\noindent \textbf{[Assumption 4]~} The matrix $(\mathds{I} - 2\theta \widehat{\mathbb{G}}^{\top})$ is invertible, and every component of the associated vector $\widehat{{\bm{\mathcal{I}}}} = (\mathds{I} - 2 \theta \widehat{\mathbb{G}}^{\top})^{-1} \cdot \mathbf{1}$ is non-negative.\\
	
When assumptions 3 and 4 hold, the unique vector of equilibrium probabilities of the legislators voting in favor of the interest group conditional on an arbitrary investment vector $\mathbf{m}$ is
\begin{align}
\begin{aligned}
\mathbf{q}^{\ast}(\mathbf{m}) ~&=~ \frac{1}{2}\cdot \mathbf{1} ~+~ \theta \cdot (\mathds{I} - 2 \theta 
\widehat{\mathbb{G}})^{-1} \cdot \big( \mathbf{u(m)} + \bm{\sigma} \big),
\end{aligned}
\end{align}
	
\noindent where the vector $\mathbf{u(m)}$ specifies resource-utilities of legislators, and the vector $\bm{\sigma}$ corresponds to the ideological inclinations of legislators. Thus, the expected vote share in favor of the interest group at any arbitrary investment vector ${\bf m}$ is 
\begin{align}
\mathcal{Q}^*(\mathbf{m}) ~=~ \big(\mathbf{q}^*({\bf m})\big)^{\top} \cdot \mathbf{1} ~=~ \frac{n}{2} ~+~ \theta \big(\mathbf{u}({\bf m})  ~+~ \bm{\sigma}\big)^{\top} \cdot \bm{\mathcal{I}^{\alpha}},
\end{align}
	
\noindent where the vector $\bm{\mathcal{I}^{\alpha}} =  (\mathds{I} - 2\theta \widehat{\mathbb{G}}^{\top})^{-1} \cdot \mathbf{1}$, and its entries denote the influence of each legislator in the presence of affective polarization between parties.
	
Let $I^{0}_{iP}$ denote the influence of a legislator $i$ in party $P$ in the \textit{absence} of affective polarization (i.e., when $\alpha = 0$). The influence of party $P$ in the absence of affective polarization will thus be $I^{0}_{P} = \sum\limits_{i \in P} I^0_{iP}$. Similarly, let $I^{\alpha}_{iP}$ and $I^{\alpha}_{P}$ respectively denote the influence of legislator $i$ belonging to party $P$ and the influence of party $P$ in the presence of affective polarization (i.e., when $\alpha >0$). We refer to influence in the absence (presence) of affective polarization as \textit{unmodified} (\textit{modified}) influence. As we show in the Appendix, 
\begin{align}
I^{\alpha}_{iP} ~=~ \big(\frac{1-\tilde{\alpha} I^0_{P'}}{1 - \tilde{\alpha}^{2} I^0_{P} I^0_{P'}}\big) \cdot I^0_{iP} ~=~ \omega_{P}  \cdot I^0_{iP},
\end{align}
	
\noindent where $\tilde{\alpha} = 2 \theta \alpha$, and the factor $\omega_{P}$ indicates how affective polarization modifies the influence a legislator has in the absence of affective polarization. The modified influence of every legislator is well-defined and strictly positive if $\tilde{\alpha} < \tilde{\alpha}_{max}$, where $\tilde{\alpha}_{max} = \min\{\frac{1}{\mathit{I}^0_{F}}, \frac{1}{\mathit{I}^0_{A}}\}$. When the modified influence of every agent is strictly positive, then investments towards every agent serve as a ``good" from the perspective of the interest group. We shall henceforth assume $\alpha < \hat{\alpha} = \frac{\tilde{\alpha}_{max}} {2 \theta}$.\\
	
%%%%%%%%%%%%%%%
%%  PROPOSITION 4     %%
%%%%%%%%%%%%%%%
	
\noindent {\bf Proposition 4.} \textit{There exists a unique equilibrium, $\big(\mathbf{m}^{\ast}, \mathbf{q}^{\ast}(\mathbf{m}^{\ast})\big)$, where the equilibrium investment vector maximizes the sum of modified influence weighted resource-utilities of legislators.}\\
	
\noindent {\bf Proof.} Using equation (23), the expected vote share in favor of the interest group at any arbitrary investment vector is
\begin{align*}
\mathcal{Q}(\mathbf{m}) ~=~  \frac{n}{2} ~+~ \theta \cdot \Big(\sum\limits_{i \in F} I^{\alpha}_{iF} \cdot u(m_{i}) ~+~ \sum\limits_{i \in A} I^{\alpha}_{iA} \cdot u(m_{i}) \Big) ~+~ \theta \sigma \cdot (I^{\alpha}_F - I^{\alpha}_A)
\end{align*} 
	
\noindent The impact of affective polarization on the expected vote share operates via how it modifies the influence of legislators. Influence, however, is independent of investments by the interest group. Hence, the interest group effectively chooses the investment vector that maximizes a weighted sum of legislators' resource-utilities, where the weights correspond to the modified influence of legislators. Weierstrass theorem ensures the existence of a maximizer and strict concavity of $Q(\mathbf{m})$ in $\mathbf{m}$ ensures uniqueness. $~\blacksquare$\\
	
%%%%%%%%%%
%%%%%%%%%%
	
To assess the impact of marginal changes in either type of polarization on the equilibrium expected vote share it is useful to note that influence of legislators is independent of the level of ideological polarization. In contrast, affective polarization modifies the influence of legislators in two ways (see equation 25). The denominator of $\omega_{P}$ highlights that the mutuality inherent in affective polarization \textit{amplifies} the unmodified influence of every legislator in both parties. Higher affective-disutility and greater unmodified influence of either party increase this amplification factor. The term in the numerator of $\omega_{P}$ \textit{dampens} the unmodified influence, where the dampening factor increases with increasing affective polarization and the unmodified influence of the opposite party. Further, the amplifying factor is identical for every legislator in both parties and decreases non-linearly with an increase in affective polarization. In contrast, the dampening factor is relatively stronger for the party with relatively lower unmodified influence and declines linearly with an increase in affective polarization. Under our assumptions, the dampening effect of affective polarization on the influence of a legislator outweighs its amplifying effect. Consequently, the influence of every legislator in the presence of affective polarization is (weakly or strictly) lower than in the absence of affective polarization.\\ 
	
%%%%%%%%%%%%%%%
%%  LEMMA 2      %%
%%%%%%%%%%%%%%%
	
\noindent {\bf Lemma 2.} \textit{Suppose the within-party networks are such that $I^0_{P} > I^0_{P'}$.}\\
\noindent {\bf (a)} \textit{The modified influence of every legislator in party $P$ decreases till affective polarization reaches a threshold, and then increases with increasing affective polarization.}\\
\noindent {\bf (b)} \textit{The modified influence of every legislator in party $P'$ monotonically decreases with an increase in affective polarization.}\\
\noindent {\bf (c)}  \textit{The difference in modified influence of the parties monotonically increases with an increase in affective polarization.}
	
\noindent {\bf Proof.} In the Appendix. $~\blacksquare$\\
	
%%%%%%%%%
%%%%%%%%%
	
The first two parts of Lemma 2 highlight that the impact of increasing affective polarization on the influence of a legislator can be monotonic or non-monotonic depending on the party the legislator belongs to. The last part essentially suggests that the party that is relatively more influential in the absence of affective polarization becomes even more so with an increase in the level of affective polarization. Formally, 
\begin{align*}
\Delta I^{\alpha} ~\equiv~ I^{\alpha}_P ~-~ I^{\alpha}_{P'} ~=~  \sum_{i \in P} I^{\alpha}_{iP} ~-~ \sum_{i \in P'} I^{\alpha}_{iP'} ~=~ \big(\frac{I^0_P - I^0_{P'}}{1 -\tilde{\alpha}^{2} I^0_{P} I^0_{P'}} \big) ~>~ I^0_P - I^0_{P'} ~\equiv~ \Delta I^{0}.
\end{align*} 
	
The relevant question is whether and when affective polarization benefits the interest group relative to the benchmark scenario with no affective polarization between the parties. In equilibrium, the expected vote share in favor of the interest group is 
\begin{align}
\mathcal{Q}^{\ast} ~=~  \frac{n}{2} ~+~ \theta \cdot \Big(\sum\limits_{i \in F} I^{\alpha}_{iF} \cdot u(m_{i}^*) ~+~ \sum\limits_{i \in A} I^{\alpha}_{iA} \cdot u(m_{i}^*) \Big) ~+~ \theta \sigma \cdot \big( \frac{I^0_F - I^0_A}{1 -\tilde{\alpha}^{2} I^0_{F} I^0_{A}} \big)
\end{align} 
	
\noindent where the equilibrium investment $m_{i}^*$ toward any legislator $i$ depends on affective polarization but not ideological polarization. Formally, the change in equilibrium expected vote share at some affective polarization $\alpha > 0$ relative to the absence of affective polarization (i.e., $\alpha = 0$) is 
\begin{align}
\Delta \mathcal{Q}^{\ast}(\alpha, 0) ~\equiv~ \mathcal{Q}^{\ast}(\alpha) ~-~ \mathcal{Q}^{\ast}(0),
\end{align}
	
\noindent The following proposition highlights the marginal impacts of the two types of polarization on the interest group and the conditions under which affective polarization can benefit the interest group.\\
	
%%%%%%%%%%%%%%%
%%  PROPOSITION 5      %%
%%%%%%%%%%%%%%%
	
\noindent {\bf Proposition 5.} \textit{Consider any network in party $F$ and any network in party $A$.}\\
\smallskip
\noindent {\bf (a)} \textit{Increase in ideological polarization strictly benefits the interest group if and only if $I^{0}_{F} > I^{0}_{A}$.}\\
\smallskip
\noindent {\bf (b)} \textit{Increase in affective polarization strictly increases the marginal benefit to the interest group from increasing ideological polarization if and only if $I^{0}_{F} > I^{0}_{A}$.}\\
\smallskip
\noindent {\bf (c)} \textit{Sufficiently small levels of affective polarization strictly hurt the interest group relative to the absence of affective polarization.}\\
\smallskip 
\noindent {\bf (d)} \textit{There exist some sufficiently high levels of affective polarization that strictly benefit the interest group relative to the absence of affective polarization if}
\begin{itemize}
\item \textit{$I^0_{F} > I^0_{A}$ and ideological polarization is sufficiently high; or,} 
\item \textit{$I^0_{A} > I^0_{F}$ and ideological polarization is sufficiently low.}
\end{itemize}
	
\noindent {\bf Proof.} In the Appendix. $~\blacksquare$\\
	
%%%%%%%%%
%%%%%%%%%
	
Part (a) highlights that, as in the baseline model, the interest group benefits from an increase in ideological polarization between parties if and only if the party in favor of the interest group has relatively more influence in the absence of affective polarization. Formally, $\frac{\partial \mathcal{Q}^{\ast}}{\partial \sigma} > 0$ if and only if $I^{0}_{F} > I^{0}_{A}$. Further, this result does not depend on the level of affective polarization. Part (b) highlights the interaction between ideological and affective polarization. As mentioned earlier, affective polarization effectively amplifies the difference in the influence of the two parties in the absence of affective polarization. Consequently, an increase in affective polarization amplifies the returns to the interest group from an increase in ideological polarization, and its qualitative sign depends on which party has more influence in the absence of affective polarization. Formally, $\frac{\partial}{\partial \alpha} \Big[\frac{\partial \mathcal{Q}^{\ast}}{\partial \sigma}\Big] > 0$ if and only if $I^{0}_{F} > I^{0}_{A}$.
	
Part (c) follows from Lemma 2(a) which shows the influence of every legislator decreases as affective polarization increases from zero up to a threshold. Sufficiently small levels of affective polarization, therefore, reduce the effectiveness of investments by the interest group, and thus hurt the interest group (relative to the absence of affective polarization). Part (d), which relies on every part of Lemma 2, essentially highlights that sufficiently high ideological and affective polarization benefits the interest group if and only if the party in favor of the interest group is relatively more influential in the absence of affective polarization.\\  
	
%%%%%%%%%%%%%%%
%%  CONCLUSION      %%
%%%%%%%%%%%%%%%
	
\noindent \textbf{\Large 5. Conclusion}\\
	
\noindent The role of special interest money in political decision-making and the impact of rising polarization among political elites have been salient themes in public discourse and academic inquiry. We investigate when can an interest group exploit increasing polarization between political parties to its advantage using monetary investments toward legislators. We consider both ideological and affective polarization between parties, where ideological polarization reflects differences in policy preferences whereas affective polarization reflects animus towards legislators belonging to the other party. The key feature of our model is that it conceives legislators as social agents and accounts for the inter-personal relations between legislators within and across parties. The relations between legislators are formalized as a network which, in turn, determines the structural influence of every legislator, and consequently the influence of each party. The influence of legislators (and parties) is independent of ideological polarization. In contrast, affective polarization effectively creates negative cross-party linkages and consequently modifies the influence of legislators. 
	
Our analysis underscores that the impact of rising ideological or affective polarization between political parties on the interest group depends crucially on the relative influence of the parties. An increase in ideological polarization benefits the interest group if and only if the party in favor of the interest group is relatively more influential. An increase in affective polarization increases the returns to the interest group from increasing ideological polarization if and only if the party in favor of the interest group in the absence of affective polarization is relatively more influential. Sufficiently small levels of affective polarization reduce the effectiveness of investments by the interest group, and thus hurt the interest group relative to the absence of affective polarization. In contrast, sufficiently high levels of ideological and affective polarization benefit the interest group if and only if the party in favor of the interest group is relatively more influential.
	
Our analysis can be extended to account for other relevant considerations (e.g., within-party ideological heterogeneity). Exploring the co-evolution of political polarization and influence activities may be particularly insightful. Our model may also be modified to investigate other questions related to polarization. For instance, there is growing suspicion that social media platforms facilitate ideological and affective polarization among the general public. Our model can serve as the starting point to investigate these issues by replacing the interest group with a firm, and appropriately modeling its action space and objective function.\\
	
%%%%%%%%%%%%
%% REFERENCES %%
%%%%%%%%%%%%
\nocite{*}
\printbibliography[title=References, type=article]

%\bibliographystyle{elsarticle-num} 
%\bibliography{Lobby}

%\include{ICPL_Jul23.bbl}
	
%%%%%%%%
%%%%%%%%
	
\newpage
	
%%%%%%%%%%%
%% APPENDIX %%
%%%%%%%%%%%

\begin{center}
{\large \bf APPENDIX}\\	
\end{center}

\bigskip
	
\noindent {\bf Expected vote share}\\	
\noindent A voting rule that would justify using $\mathcal{Q}$ -- expected vote share -- as the appropriate objective function for the interest group is \textit{Random Dictatorship}. Under Random Dictatorship, each legislator is equally likely to be selected as the dictator, and the dictator's vote determines the outcome. If $q_i$ is the probability that legislator $i$ votes for policy $f$, then the ex-ante probability that policy $f$ is the outcome under random dictatorship will be $\tilde{\mathcal{Q}} = \frac{1}{n} \sum_{i \in L} q_i$. As expected vote share $\mathcal{Q} ~=~ n \cdot \tilde{\mathcal{Q}}$, maximizing $\mathcal{Q}$ is equivalent to maximizing $\tilde{\mathcal{Q}}$. Hence, $\mathcal{Q}$ is technically justifiable if the outcome is determined by random dictatorship.
	
A related question is: when is $\mathcal{Q}$ justifiable under majority voting? The answer lies in identifying the conditions that imply the probability of each legislator being \textit{pivotal} is identical under both majority voting and random dictatorship. Each legislator is pivotal with probability $\frac{1}{n}$ under random dictatorship. Under majority voting, each legislator is pivotal with probability $\frac{1}{n}$ if and only if every legislator is equally likely to vote for both policies. Consider, for example, a setting with three legislators. Let $q_i$ be the probability that legislator $i \in \{1,2,3\}$ votes for policy $f$. The probability that a legislator $i$ is pivotal -- $\pi^{mv}_i$ -- under majority voting is the probability that the remaining two legislators do not vote for the same policy. Hence, 
$$\pi^{mv}_i ~=~ q_j (1- q_k) ~+~ q_k (1- q_j), ~~~~\mbox{where}~j \ne i \ne k$$
	
\noindent The unique solution to this system of three equations in three unknowns is $q_i = \frac{1}{2}$ for every $i \in \{1,2,3\}$. Thus, $\mathcal{Q}$ justifiable under majority voting if every legislator is equally likely to vote for both policies. Hence, our results are most informative in contexts where every legislator in ``on the fence".\\
	
%%%%%%%%%%%%%%%
%%  ASSUMPTION 1      %%
%%%%%%%%%%%%%%%
	
\noindent \textbf{Discussion of Assumption 1}\\ 
\noindent Let $\bar{q}(P)$ and $\underbar{q}(P)$ denote the highest and lowest probabilities of any legislator in party $P $ voting for policy-$f$. The highest probability of any legislator in party $P$, say, legislator $i$, voting in favor of the interest group is
\begin{align}
q_{i}^{\max}(P) ~=~ \min\{1, \bar{q}(P)\},  
\end{align}

\noindent where, if $\bar{q}(P) \in [0,1]$, then it would be given by 
\begin{align}
\bar{q}(P) ~=~ \frac{1}{2} ~+~ \theta \Big(u(M)~+~ \delta~(n - 1) ~+~ \sigma_{P} \Big)
\end{align}

\noindent This is because $\bar{q}(P)$ would correspond to the scenario where (i) the interest group invests all its resources on legislator $i$, (ii) $g_{ij} =1$ for all $j \ne i$, and (iii) $q_{j} = 1$ for all $j \ne i$. Analogously, the lowest probability of a legislator in party $P$, say, legislator $i$, voting in favor of the interest group is
\begin{align}
q_{i}^{\min}(P) ~&=~ \max\{0, \underbar{q}(P)\},
\end{align}

\noindent where, if $\underbar{q}(P) \in [0,1]$, then it would be given by 
\begin{align}
\underbar{q}(P) ~&=~ \frac{1}{2} ~+~ \theta \Big(- \delta~(n - 1) ~+~ \sigma_P \Big)
\end{align}
	
\bigskip
	
%%%%%%%%%%%%%%%
%%  LEMMA 1      %%
%%%%%%%%%%%%%%%
	
\noindent \textbf{Proof of Lemma 1}\\
\noindent Consider a pair of networks $\mathbb{G}$ and $\mathbb{G}^{+} = \mathbb{G} + \{\vec{ij} \}$. Define $\Delta{\mathbb{G}} = (\mathbb{G}^{+} - \mathbb{G})$, wherein the entry in the $i$-th row and $j$-th column is 1 and zero elsewhere. The influence vector under network $\mathbb{G}^{+}$ is 
\begin{align}
\bm{\mathcal{I}}^{+} ~=~ (\mathds{I} - \beta \cdot (\mathbb{G}^{+})^{\top})^{-1} \cdot \mathbf{1} ~=~ \big[~(\mathds{I} - \beta 
\mathbb{G}^{\top}) - \beta (\Delta \mathbb{G})^{\top}  ~\big]^{-1} \cdot \mathbf{1}
\end{align}
\noindent Also, 
\begin{align}
(\Delta \mathbb{G})^{\top} ~&=~ (\mathbb{G}^{+} - \mathbb{G})^{\top}  ~=~
\begin{bmatrix}
0 &  \cdots & 0 \\
\vdots & \cdots ~ g_{ji} ~ \cdots & \vdots \\
0 & \cdots & 0
\end{bmatrix} 
~=~ g_{ji} \cdot \begin{bmatrix}
0 &  \cdots & 0 \\
\vdots & \cdots ~ 1_{ji} ~ \cdots & \vdots \\
0 & \cdots & 0
\end{bmatrix} ~=~ g_{ji} \cdot \mathbb{E} 
\end{align}
\noindent where $\mathbb{E}$ is a matrix with 1 in $j$-th row and $i$-th column and zero elsewhere. The matrix $\mathbb{E}$ can be represented as a product of the $n \times 1$ column vector $\mathbf{y} = (0, \ldots, 1_{j}, \ldots, 0)^{\top}$ and the $1 \times n$ row vector $\mathbf{z} = (0, \ldots, 1_{i}, \ldots, 0)$. Hence, using the Sherman-Morrison formula,
\begin{align}
\bm{\mathcal{I}}^{+} ~&=~ \big[~(\mathds{I} - \beta \mathbb{G}^{\top}) - \beta g_{ji} \cdot \mathbf{y} \cdot \mathbf{z}~\big]^{-1} \cdot \mathbf{1} ~=~ (\mathds{I} - \beta \mathbb{G}^{\top})^{-1} \cdot \mathbf{1} ~+~ \frac{\beta \cdot 
\mathit{I}_{i}}{1 - \beta \cdot x_{ij}} \cdot \big(x_{1j} , ~ \cdots~ , x_{nj} \big)^{\top}\\ \notag
~&=~ \bm{\mathcal{I}} ~+~ \big( \frac{\beta \cdot \mathit{I}_{i}}{1 - \beta \cdot x_{ij}} \big) \cdot \big(x_{1j} , ~ \cdots ~, x_{nj} \big)^{\top}
\end{align}
\noindent As described in Section 3.1, $I_{i} \ge 1$, $\beta > 0$, and every $x_{ij} \ge 0$. Hence, all we need to show is that the term $(1 - \beta ~x_{ij})$ is strictly positive. An upper bound for $x_{ij}$ can be found by considering the complete network, $\mathbb{G}^c$. Further, $(\mathbb{G}^c)^{\top} = \mathbb{G}^c$. Now, let the diagonal entries of the matrix $(\mathds{I} - \beta \mathbb{G}^c)^{-1}$ be denoted by $x_d^{c}$. Similarly, let the off-diagonal entries be denoted by $x_{od}^{c}$. Algebraic calculations suggest 
\begin{align}
x_{d}^{c} ~=~ \frac{1 - (n-2) \beta}{1 - (n-2) \beta - (n-1) \beta^{2}} ~~~\mbox{and}~~~ x_{od}^{c} ~=~ \frac{\beta}{1 - (n-2) 
\beta - (n-1) \beta^{2}}
\end{align}
	
Using the above expressions in equation (33) it can be shown that $(1 - \beta ~x_{ii}^{c}) > 0$ and $(1 - \beta ~x_{ij}^{c}) > 0$ if and only if $\beta < \frac{1}{n}$. The sum of entries in each row of $\mathbb{G}^c$ is $(n-1)$. Hence, Perron-Frobenius theorem implies the largest eigenvalue of $\mathbb{G}^c$ is $(n-1)$. A sufficient condition for the invertibility of $(\mathds{I} - \beta \mathbb{G}^c)$, and consequently for the influence vector to be well defined, is that $\beta < \frac{1}{n-1}$. Consequently, $\beta < \frac{1}{n}$ not only ensures the influence vector is well-defined, but also that influence of every legislator will increase with the addition of any (feasible) new directed link in any given network. $~\blacksquare$\\
	
%%%%%%%%%%%%%%%%%%%
%%  PROOF PROPOSITION 3     %%
%%%%%%%%%%%%%%%%%%%
	
\noindent \textbf{Proof of Proposition 3}\\ 
\noindent Consider a pair of networks $\mathbb{G}$ and $\mathbb{G}^{+} = \mathbb{G} + \{\vec{ij}\}$ such that $\mathbb{G}^+$ is relatively stronger. Lemma 1 implies $I^{+}_i \geq I_i$ for every legislator $i \in L$. Let $\mathbf{m}^{\ast}$ and $\mathbf{m}^{\ast +}$ denote the equilibrium investment vectors under $\mathbb{G}$ and $\mathbb{G}^{+}$, respectively.
This change in network benefits the interest group if 
\begin{align}
\mathcal{Q}^*(\mathbb{G}^+) ~-~ \mathcal{Q}^*(\mathbb{G}) ~=~ \big(\sum\limits_{i \in L} \mathit{I}^{+}_{i} \cdot u(m_{i}^{+ \ast}) ~-~ \sum\limits_{i \in L} \mathit{I}_{i} \cdot u(m_{i}^{\ast}) \big) ~+~ \sigma \cdot \big(\Delta I_F ~-~ \Delta I_A \big) ~\geq 0,	
\end{align}
	
\noindent We first note that, by definition,
\begin{align}
& \mathcal{Q}^{\ast}(\mathbf{m}^{\ast +}, \mathbb{G}^{+}) ~\equiv~ \max\limits_{\mathbf{m}} \mathcal{Q}(\mathbf{m}, \mathbb{G}^{+}) ~\geq~ \mathcal{Q}(\mathbf{m}^{\ast}, \mathbb{G}^{+}) \\
\implies ~~ & \sum\limits_{i} u(m_{i}^{\ast +}) \cdot \mathit{I}^{+}_{i} ~\geq~ \sum\limits_{i} u(m_{i}^{\ast} ) \cdot \mathit{I}_{i}^{+}\\
\implies ~~ & \sum\limits_{i} u(m_{i}^{\ast +}) \cdot \mathit{I}^{+}_{i} ~\geq~ \sum\limits_{i} u(m_{i}^{\ast} ) \cdot \mathit{I}_{i}
\end{align}
	
\noindent Thus, the investment effect is always non-negative. If  $\Delta I_{F} \geq \Delta I_{A}$, then the polarization effect is also non-negative. Consequently, the interest group will weakly or strictly gain from the change in the network. In contrast, if $\Delta I_{A} > \Delta I_{F}$, then the polarization effect is strictly negative. The interest group will benefit from the change in network only when ideological polarization is below a certain threshold $\hat{\sigma}$, where 
\begin{align}
\hat{\sigma} ~ = ~ \frac{\sum\limits_{i} ~I_i^+ \cdot u(m_{i}^{+ \ast}) ~-~ \sum\limits_{i} ~I_i \cdot u(m_{i}^{\ast})}{\Delta I_{A}  - \Delta I_{F}} ~>~ 0 
\end{align} 

\bigskip
%%%%%%%%%%%%%%%%%%%%%%%%%%%%%
%%%%%%%%%%%%%%%%%%%%%%%%%%%%%
	
\noindent \textbf{\large Influence under affective polarization}\\
\noindent We begin with some further remarks about assumption A4. The matrix $(\mathds{I} - 2 \theta \widehat{\mathbb{G}})$ must be invertible for $\mathbf{q}^{\ast}(\mathbf{m})$ -- the equilibrium voting probabilities conditional on some investment vector $\bf{m}$ -- to be well defined. In general, the inverse of a $2 \times 2$ block matrix is given by
\begin{align}
\begin{aligned}
\left[ 
\begin{array}{c|c} 
\mathbf{W} & \mathbf{X} \\ 
\hline
\mathbf{Y} & \mathbf{Z} \\
\end{array} \right]^{-1} ~=~ \left[
\begin{array}{c|c}
(\mathbf{W} ~-~ \mathbf{X} \mathbf{Z}^{-1}\mathbf{Y})^{-1} & - (\mathbf{W} ~-~ \mathbf{X} \mathbf{Z}^{-1}\mathbf{Y})^{-1}\mathbf{X} \mathbf{Z}^{-1} \\
\hline
- (\mathbf{Z} ~-~ \mathbf{Y} \mathbf{W}^{-1}\mathbf{X})^{-1} \mathbf{Y} \mathbf{W}^{-1} & (\mathbf{Z} ~-~ \mathbf{Y} \mathbf{W}^{-1}\mathbf{X})^{-1} \\
\end{array} \right]
\end{aligned}
\end{align}
\noindent where $\mathbf{W}$ and $\mathbf{Z}$ are square matrices and $\mathbf{X}$ and $\mathbf{Y}$ are conformable for partition (see Bernstein, 2018). For the inverse to exist, the matrices $\mathbf{W}$, $\mathbf{Z}$, $(\mathbf{W} ~-~ \mathbf{X} \mathbf{Z}^{-1}\mathbf{Y})$ and $(\mathbf{Z} ~-~ \mathbf{Y} \mathbf{W}^{-1}\mathbf{X})$ must themselves be invertible. 
	
In our model, 
\begin{align}
\begin{aligned}
&\quad ( \mathds{I} - 2\theta \widehat{\mathbb{G}}^{\top})^{-1} ~=~ \left[\begin{array}{c|c} (\mathds{I} - \beta \mathbb{G}_{FF}^{\top}) & \tilde{\alpha} \cdot \mathds{1}_{AF}^{\top}\\ 
\hline
\tilde{\alpha} \cdot \mathds{1}_{FA}^{\top} & ( \mathds{I} - \beta \mathbb{G}_{AA}^{\top}) \\
\end{array} \right]^{-1} ~~~~~~~~ [~\because ~ \mathds{1}_{FA}^{\top} = \mathds{1}_{AF} \text{~and~} 
\mathds{1}_{AF}^{\top} = \mathds{1}_{FA} ~]\\
%&~~ \\
&= ~ \footnotesize{\left[ 
\begin{array}{c|c} 
( \mathds{I} ~-~ \beta \mathbb{G}_{FF} - \tilde{\alpha}^{2} {I}_{A}^{0} \mathds{1}_{FF})^{-1} & -~ \tilde{\alpha} ( \mathds{I} - \beta \mathbb{G}_{FF}^{\top} ~-~ \tilde{\alpha}^{2}{I}_{A}^{0} \mathds{1}_{FF} )^{-1} \mathds{1}_{FA} ( \mathds{I} - \beta \mathbb{G}_{AA}^{\top})^{-1} \\ 
\hline
-~ \tilde{\alpha}( \mathds{I} - \beta \mathbb{G}_{AA}^{\top} ~-~ \tilde{\alpha}^{2} {I}_{F}^{0} \mathds{1}_{AA} )^{-1} \mathds{1}_{AF} ( \mathds{I} - \beta \mathbb{G}_{FF}^{\top})^{-1} & ( \mathds{I} - \beta \mathbb{G}_{AA}^{\top} ~-~ \tilde{\alpha}^{2} {I}_{F}^{0} \mathds{1}_{AA} )^{-1} \\
\end{array} \right]}\\
\end{aligned}
\end{align}
	
\noindent Hence, the matrix $(\mathds{I} - 2\theta \widehat{\mathbb{G}}^{\top})$ is invertible if and only if the matrices $(\mathds{I} - \beta \mathbb{G}_{PP}^{\top})$ and $ (\mathds{I} - \beta \mathbb{G}_{PP}^{\top} - \tilde{\alpha}^{2}{I}_{P'}^{0} \mathds{1})$ are invertible for $P, P' \in \{F, A\}$ and $P \neq P'$. Further, when $(\mathds{I} - 2\theta \mathbb{G}^{\top})$ is invertible, then the modified influence vector is 
\begin{align*}
\begin{aligned}
{\bm{\mathcal{I}}^{\alpha}} ~=~  ( \mathds{I} ~-~ 2\theta \widehat{ \mathbb{G}}^{\top})^{-1}\cdot \mathbf{1} ~=~ \left[ 
\begin{array}{c|c} 
( \mathds{I} ~-~ \beta \mathbb{G}_{FF}^{\top}) & \tilde{\alpha} \cdot \mathds{1}_{FA}\\
\hline
\tilde{\alpha} \cdot \mathds{1}_{AF} & ( \mathds{I} ~-~ \beta \mathbb{G}_{AA}) \\
\end{array} \right]^{-1} \cdot \mathbf{1} ~~~~~~~~~~~~~~~~~~~~~~~~~~~~~~  \\
~~ \\
=  ~ \footnotesize{\left[ 
\begin{array}{c|c} 
( \mathds{I} ~-~ \beta \mathbb{G}_{FF} - \tilde{\alpha}^{2} {I}_{A}^{0} \mathds{1}_{FF})^{-1} & -~ \tilde{\alpha} ( \mathds{I} - \beta \mathbb{G}_{FF}^{\top} ~-~ \tilde{\alpha}^{2} {I}_{A}^{0} \mathds{1}_{FF} )^{-1} \mathds{1}_{FA} ( \mathds{I} - \beta \mathbb{G}_{AA}^{\top})^{-1} \\ 
\hline
-~ \tilde{\alpha}( \mathds{I} - \beta \mathbb{G}_{AA}^{\top} ~-~ \tilde{\alpha}^{2} {I}_{F}^{0} \mathds{1}_{AA} )^{-1} \mathds{1}_{AF} ( \mathds{I} - \beta \mathbb{G}_{FF}^{\top})^{-1} & ( \mathds{I} - \beta \mathbb{G}_{AA}^{\top} ~-~ \tilde{\alpha}^{2} {I}_{F}^{0} \mathds{1}_{AA} )^{-1} \\
\end{array} \right]}\cdot \mathbf{1} \qquad \\
~~\\
= ~ \footnotesize{\left[ 
\begin{array}{c c} 
\mathds{I} ~-~ \beta \mathbb{G}_{FF} - \tilde{\alpha}^{2} {I}_{A}^{0} \mathds{1}_{FF})^{-1} \cdot \mathbf{1}_{(n_{F} \times 1)} ~-~ \tilde{\alpha} ( \mathds{I} - \beta \mathbb{G}_{FF}^{\top} ~-~ \tilde{\alpha}^{2} {I}_{A}^{0} \mathds{1}_{FF} )^{-1} \mathds{1}_{FA} ( \mathds{I} - \beta \mathbb{G}_{AA}^{\top})^{-1}  \cdot \mathbf{1}_{(n_{A} \times 1)}  \\
~& \\
-~ \tilde{\alpha}( \mathds{I} - \beta \mathbb{G}_{AA}^{\top} ~-~ \tilde{\alpha}^{2} {I}_{F}^{0} \mathds{1}_{AA} )^{-1} \mathds{1}_{AF} ( \mathds{I} - \beta \mathbb{G}_{FF}^{\top})^{-1}\cdot \mathbf{1}_{(n_{F} \times 1)} ~+~ ( \mathds{I} - \beta \mathbb{G}_{AA}^{\top} ~-~ \tilde{\alpha}^{2} {I}_{F}^{0} \mathds{1}_{AA} )^{-1}\cdot \mathbf{1}_{(n_{A} \times 1)} \\
\end{array} \right]} \qquad ~~~
\end{aligned}
\end{align*}
	
\begin{align}
\begin{aligned}
&= ~ \normalsize{\left[ 
\begin{array}{c c} 
\big( \frac{1}{1 - \tilde{\alpha}^{2} \cdot {I}_{A}^{0} {I}_{F}^{0}} \big) \cdot \bm{\mathcal{I}}_{F}^{0} ~-~\big( \frac{\tilde{\alpha} {I}_{A}^{0}}{1 - \tilde{\alpha}^{2} \cdot {I}_{A}^{0} {I}_{F}^{0}} \big) \cdot \bm{\mathcal{I}}_{F}^{0}  \\
~ \\
-~\big( \frac{\tilde{\alpha} {I}_{F}^{0}}{1 - \tilde{\alpha}^{2} \cdot {I}_{A}^{0} {I}_{F}^{0}} \big) \cdot \bm{\mathcal{I}}_{A}^{0} ~+~ \big( \frac{1}{1 - \tilde{\alpha}^{2} \cdot {I}_{A}^{0} {I}_{F}^{0}} \big) \cdot \bm{\mathcal{I}}_{A}^{0}  \\
\end{array} \right]} ~~~~~~ \\
&~~ \\
~&=~ \normalsize{\left[ 
\begin{array}{c c} 
\omega_{F} \cdot \bm{\mathcal{I}}_{F}^{0}  \\
\omega_{A} \cdot \bm{\mathcal{I}}_{A}^{0}  \\
\end{array} \right]}
\end{aligned}
\end{align}
\noindent where $\bm{\mathcal{I}}_{P}^{0}$ is the $n_{P} \times 1$ influence vector of legislators in party $P \in \{F, A\}$ in the absence of affective polarization.\\
	
%%%%%%%%%%%%%%%%%%%
%%  PROOF of  LEMMA 2      %%
%%%%%%%%%%%%%%%%%%%
	
\noindent {\bf Proof of Lemma 2}\\ 
\noindent Suppose party $P$ is more influential than party $P'$, i.e. $I_{P}^{0} > I_{P'}^{0}$. It follows that  
\begin{align}
\omega_{P}(\alpha) ~=~ \frac{(1-\tilde{\alpha}I_{P'}^{0})}{(1-\tilde{\alpha}^{2}I_{P}^{0} I_{P'}^{0})} ~>~ \frac{(1-\tilde{\alpha}I_{P}^{0})}{(1-\tilde{\alpha}^{2} I_{P}^{0} I_{P'}^{0})} ~=~ \omega_{P'}(\alpha)  
\end{align}
where $\omega_{P}(\alpha)$ is the factor that modifies the influence of any legislator in party $P$. The above suggests 
\begin{itemize}
\item $\omega_{P}(0) ~=~ \omega_{P'}(0) ~=~ 1$
\item $\lim\limits_{\alpha \rightarrow \hat{\alpha}} \omega_{P} ~=~ 1$
\item $\lim\limits_{\alpha \rightarrow \hat{\alpha}} \omega_{P'} ~=~ 0$
\end{itemize} 
	
To prove part (a), we first note that
\begin{align}
\frac{\partial ~ \omega_{P}}{\partial \alpha} ~&=~ \frac{-2 \theta I_{P'}^{0}}{(1 - \tilde{\alpha}^{2}I_{P}^{0} I_{P'}^{0})^{2}}~\big[1 ~+~ \tilde{\alpha}^{2} I_{P}^{0} I_{P'}^{0} ~-~ 2\tilde{\alpha} I_{P}^{0} \big] 
\end{align}
	
\noindent Thus, if $\omega_{P} > \omega_{P'}$, then $\omega_{P}$ attains a minimum at $\alpha^* = \frac{1}{2 \theta I_{P'}^{0}}\big( 1 - \sqrt{1 - \frac{I_{P'}^{0}} {I_{P}^{0}}} \big)$. Consequently, the modified influence of every legislator in party $P$ decreases till affective polarization reaches the threshold value of $\alpha^*$, and then increases with increasing affective polarization.
	
To prove part (b), we first note that
\begin{align}
\frac{\partial ~ \omega_{P'}}{\partial \alpha} ~&=~ \frac{-2 \theta I_{P}^{0}}{(1 - \tilde{\alpha}^{2} I_{P}^{0} I_{P'}^{0})^{2}}~\big(1 ~+~ \tilde{\alpha}^{2}I_{P}^{0} I_{P'}^{0} ~-~ 2\tilde{\alpha} I_{P'}^{0} \big) ~=~ \omega_{P} ~-~  \tilde{\alpha} I_{P'}^{0} ~ \omega_{P'}
\end{align}
	
\noindent The above derivative is strictly negative because $\omega_{P'} < \omega_{P}$ and  $\tilde{\alpha} I_{P'}^{0} \in (0,1)$ for every $\alpha \in [0, \hat{\alpha})$. Thus, the modified influence of every legislator in party $P'$ monotonically decreases with increasing affective polarization. 
	
Finally, part (c) follows from the observation that for $\alpha \in [0, \hat{\alpha})$,
\begin{align}
\frac{\partial}{\partial \alpha} \big(\Delta I^{\alpha} \big) ~=~ \frac{2 \theta \tilde{\alpha}(I^0_P - I^0_{P'})}{(1 - \tilde{\alpha}^{2} I^0_{P} I^0_{P'})^2} ~>~0. 
\end{align}
	
\bigskip

%%%%%%%%%%%%%%%%%%%
%%  PROOF of  PROPOSITION 5     %%
%%%%%%%%%%%%%%%%%%%

\noindent {\bf Proof of Proposition 5}\\ 
\noindent The equilibrium expected vote share in favor of the interest group is
\begin{align}
\mathcal{Q}^{\ast} ~=~  \frac{n}{2} ~+~ \theta \cdot \Big(\sum\limits_{i \in F} I^{\alpha}_{iF} \cdot u(m_{i}^*) ~+~ \sum\limits_{i \in A} I^{\alpha}_{iA} \cdot u(m_{i}^*) \Big) ~+~ \theta \sigma \cdot \big( \frac{I^{0}_{F} - I^{0}_{A}}{1 - \tilde{\alpha}^{2} I^0_{F} I^0_{A}} \big)
\end{align} 
	
\noindent where the equilibrium investment $m_{i}^{\ast}$ toward any legislator $i$ depends on affective polarization $\alpha$ but is independent of ideological polarization $\sigma$. Part (a) follows from the observation that $\frac{\partial \mathcal{Q}^{\ast}}{\partial \sigma} >0 $ if and only if $I_{F}^{0} > I_{A}^{0}$. Similarly, part (b) follows from the observation that $\frac{\partial}{\partial \alpha} [\frac{\partial \mathcal{Q}^{\ast}}{\partial \sigma}] >0 $ if and only if $I_{F}^{0} > I_{A}^{0}$. 
	
To prove part (c), first, recall $I^{\alpha}_{iP} ~=~ \omega_{P} {I}_{F}^{0}$. Hence, the equilibrium expected vote share can be re-written as 
\begin{align}
\begin{aligned}
\mathcal{Q}^{\ast} ~=~ \frac{n}{2} ~+~ \omega_{P} \sum\limits_{i \in P} {I}_{iP}^{0} \cdot u(m^{\ast}_{i}) ~+~ \omega_{P'}\sum\limits_{i \in P'} {I}_{iP'}^{0} \cdot u(m^{\ast}_{i}) ~+~ \frac{\theta \sigma ({I}_{F}^{0} - {I}_{A}^{0})}{(1 - \tilde{\alpha}^{2} {I}_{F}^{0} {I}_{A}^{0})}
\end{aligned}
\end{align}
	
\noindent Further, $\mathcal{Q}^{\ast}$ is continuously differentiable in $\alpha$, with
\begin{align}\notag
\frac{d \mathcal{Q}^{\ast}}{d \alpha} ~=~ \theta &\Bigg[~ \frac{\partial ~ \omega_{P}}{\partial \alpha} \cdot \sum\limits_{i \in P} {I}_{iP}^{0} \cdot u(m_{i}^{\ast}) ~+~ \omega_{P} \cdot \sum\limits_{i \in P}  {I}_{iP}^{0} \cdot u'(m_{i}^{\ast})~ \frac{\partial m_{i}^{\ast}}{\partial \alpha}\Bigg]\\ \notag 
&~+~ \theta \Bigg[~ \frac{\partial ~ \omega_{P'}}{\partial \alpha} \cdot \sum\limits_{i \in P'} {I}_{iP'}^{0} \cdot u(m_{i}^{\ast}) ~+~ \omega_{P'} \cdot \sum\limits_{i \in P'} {I}_{iP'}^{0} \cdot u'(m_{i}^{\ast})~ \frac{\partial m_{i}^{\ast}}{\partial \alpha}   \Bigg] \\
&~+~ \frac{4 \sigma \theta^{2} \cdot \tilde{\alpha}{I}_{A}^{0} {I}_{F}^{0}~({I}_{F}^{0} - {I}_{A}^{0}) }{(1 - \tilde{\alpha}^{2}{I}_{F}^{0} {I}_{A}^{0})^{2}}
\end{align}
	
\noindent The equilibrium Lagrange multiplier associated with the interest group's constraint optimization problem is $\lambda^{\ast} = \theta \cdot \omega_{P} \cdot {I}_{iP}^{0} \cdot u'(m_{i})$. Hence, the above equation can be rewritten as
\begin{align}\notag
\frac{d \mathcal{Q}^{\ast}}{d \alpha} ~=~ \theta &\Bigg[~ \frac{\partial ~ \omega_{P}}{\partial \alpha} \cdot \sum\limits_{i \in P} {I}_{iP}^{0} \cdot u(m_{i}^{\ast}) ~+~ \frac{\partial ~ \omega_{P'}}{\partial \alpha} \cdot \sum\limits_{i \in P'} {I}_{iP'}^{0} \cdot u(m_{i}^{\ast}) ~+~ \lambda^{\ast} \big( \sum\limits_{i \in P} \frac{\partial m_{i}^{\ast}}{\partial \alpha} ~+~ \sum\limits_{i \in P'}  \frac{\partial m_{i}^{\ast}}{\partial \alpha} \big)   \Bigg] \\ 
&~+~ \frac{4 \sigma \theta^{2} \cdot \tilde{\alpha} {I}_{A}^{0} {I}_{F}^{0}~({I}_{F}^{0} - {I}_{A}^{0}) }{(1 - \tilde{\alpha}^{2}{I}_{F}^{0} {I}_{A}^{0})^{2}}
\end{align}
	
\noindent As the interest group invests all of its budget in equilibrium, it must hold that $\sum\limits_{i \in P} \frac{\partial m_{i}^{\ast}}{\partial \alpha} ~+~  \sum\limits_{i\in P'}  \frac{\partial m_{i}^{\ast}}{\partial \alpha} = 0$. Hence, the previous equation becomes
\begin{align}
\frac{d \mathcal{Q}^{\ast}}{d \alpha} ~=~ \theta &\Bigg[~ \frac{\partial ~ \omega_{P}}{\partial \alpha} \cdot \sum\limits_{i \in P} {I}_{iP}^{0} \cdot u(m_{i}^{\ast}) ~+~ \frac{\partial ~ \omega_{P'}}{\partial \alpha} \cdot \sum\limits_{i \in P'} {I}_{iP'}^{0} \cdot u(m_{i}^{\ast})  \Bigg] ~+~ \frac{4 \sigma \theta^{2} \cdot \tilde{\alpha} {I}_{A}^{0} {I}_{F}^{0}~({I}_{F}^{0} - {I}_{A}^{0}) }{(1 - \tilde{\alpha}^{2}{I}_{F}^{0} {I}_{A}^{0})^{2}}
\end{align}
	
\noindent Using equation (48) and Lemma 2, $\frac{d \mathcal{Q}^{\ast}}{d \alpha} < 0$ at $\alpha = 0$. As ${I}_{P}^{0} > {I}_{P'}^{0}$, it follows that $\omega_{P'}(\hat{\alpha}) = 0$. Consequently, $m_{i}^{\ast}  \rightarrow 0$ for all $i \in P'$ when $\alpha \rightarrow \hat{\alpha}$. Hence,
\begin{align}
\lim_{\alpha \rightarrow \hat{\alpha}} \Big[\frac{d \mathcal{Q}^{\ast}}{d \alpha}\Big] ~=~  \frac{2 \theta^2 {I}_{P}^{0} {I}_{P'}^{0}}{({I}_{P}^{0} - {I}_{P'}^{0})} \cdot \sum\limits_{i \in P} {I}_{iP}^{0} \cdot u(m_{i}^{\ast}) ~+~ \frac{4 \theta^{2} \sigma \cdot {I}_{A}^{0} {I}_{F}^{0 ~2}}{({I}_{F}^{0} - {I}_{A}^{0})}
\end{align}
	
\noindent We need to consider three possible cases.\\
	
\noindent {\it Case 1: $P =F$ and $P'=A$ such that $I_{F}^{0} > I_{A}^{0}$}. In this case, both terms on the R.H.S. in the above equation are positive. Hence, $\lim\limits_{\alpha \rightarrow \hat{\alpha}} \Big[\frac{d \mathcal{Q}^{\ast}}{d \alpha}\Big] >0$. We know that $\Big[ \frac{d \mathcal{Q}^{\ast}}{d \alpha} \Big]_{\alpha = 0} < 0$. Consequently, there exists a critical level of affective polarization $\alpha_{1} \in (0, \hat{\alpha})$ such that $\Big[\frac{d \mathcal{Q}^{\ast}}{d \alpha}\Big]_{\alpha = \alpha_{1}} = 0$.\\
	
\noindent {\it Case 2: $P =A$ and $P'=F$ such that $I_{A}^{0} > I_{F}^{0}$}. We know that $\Big[ \frac{d \mathcal{Q}^{\ast}}{d \alpha} \Big]_{\alpha = 0} < 0$. Moreover, the first term in equation (49) is positive but the second term is negative. Hence the overall impact of change in $\alpha$ on $\mathcal{Q}^{\ast}$ is ambiguous. Note that, 
$$\lim\limits_{\alpha \rightarrow \hat{\alpha}} \Big[\frac{d  \mathcal{Q}^{\ast}}{d \alpha}\Big] > 0 ~~~~~ \text{if} ~~~~~ \sigma ~<~ \frac{1}{2} \cdot \Big[ \sum\limits_{i \in A} \frac{I_{iA}^{0}}{ I_{F}^{0}}~ u(m_{i}) \Big] ~\equiv~ \sigma_{1}$$
	
\noindent Hence, if $\sigma < \sigma_{1}$, then there exists a critical level of affective polarization $\alpha_{2} \in (0, \hat{\alpha})$ such that $\Big[\frac{d \mathcal{Q}^{\ast}}{d\alpha}\Big]_{\alpha = \alpha_{2}} = 0$. On the other hand, if $\sigma > \sigma_{1}$ then $\lim\limits_{\alpha \rightarrow \hat{\alpha}} \Big[\frac{d \mathcal{Q}^{\ast}}{d \alpha}\Big] < 0$.\\

\noindent {\it Case 3: $I_{F}^{0} = I_{A}^{0}$}. In this case, $\frac{\partial \omega_{P}}{\partial \alpha} < 0$ for $P \in \{F, A\}$. Hence, the first expression in equation (48) is always negative while the second term is zero. Hence, if $I_{F}^{0} = I_{A}^{0}$, then a marginal increase in affective polarization hurts the interest group.\\
	
Turning to the proof of part (d), we first define
\begin{align}
\Delta Q^{\ast}(\hat{\alpha}, 0) ~\equiv~ \mathcal{Q}^{\ast}(\alpha \rightarrow \hat{\alpha}) ~-~ \mathcal{Q}^{\ast}(\alpha = 0)
\end{align}
	
\noindent Using Lemma 2 and equation (26), if party $P$ has more influence than party $P'$ in the absence of affective polarization, then  
\begin{align}
\Delta Q^{\ast}(\hat{\alpha}, 0) ~=~\sum\limits_{i \in P} I^0_i ~ \Big(u(m_{i}^{\ast})\big|_{\alpha \rightarrow \hat{\alpha}} ~-~ u(m_{i}^{\ast})\big|_{\alpha = 0} ~\Big) ~-~ \sum\limits_{j \in P'} I^0_j ~ \Big(~ u(m_{j}^{\ast})\big|_{\alpha = 0} ~-~ \sigma_P ~ \Big),
\end{align}
\noindent where $\sigma_{P} = \sigma$ if $P = F$ and $-\sigma$ if $P = A$.\\
	
\noindent {\it Case 1: $P =F$ and $P'=A$ such that $I_{F}^{0} > I_{A}^{0}$}. Using Lemma 2 and $\mathcal{Q}^{\ast}(\alpha = 0)$, we obtain
\begin{align} \notag
\Delta Q^{\ast}(\hat{\alpha}, 0) ~=~\sum\limits_{i \in F} I^0_{iF} ~ \Big(u(m_{i}^{\ast})\big|_{\alpha \rightarrow \hat{\alpha}} ~-~ u(m_{i}^{\ast})\big|_{\alpha = 0} ~\Big) ~-~ \sum\limits_{i \in A} I^0_{iA} ~ \Big(~ u(m_{j}^{\ast})\big|_{\alpha = 0} ~-~ \sigma ~ \Big) \\ 
~=~ \sum\limits_{i \in F} I^0_{iF} ~ \Big(u(m_{i}^{\ast})\big|_{\alpha \rightarrow \hat{\alpha}} ~-~ u(m_{i}^{\ast})\big|_{\alpha = 0} ~\Big) ~-~ \sum\limits_{i \in A} I^0_{iA} ~ \Big(~ u(m_{j}^{\ast})\big|_{\alpha = 0} \big) ~+~ \sigma \cdot I_{A}^{0}
\end{align}

\noindent Hence, $\Delta Q^{\ast}(\hat{\alpha}, 0) $ is strictly positive if 
	
$$ \sigma ~>~ \sum\limits_{i \in A} \big( \frac{I_{iA}^{0}}{I_{A}^{0}} \big) \cdot \big( u(m_{j}^{\ast})\big|_{\alpha = 0} \big) - \sum\limits_{i \in F} \big( \frac{I_{iF}^{0}}{I_{A}^{0}} \big) \cdot \Big(u(m_{i}^{\ast})\big|_{\alpha \rightarrow \hat{\alpha}} ~-~ u(m_{i}^{\ast})\big|_{\alpha = 0} ~\Big) \equiv \sigma_{2}$$
	
\noindent If $\sigma > \sigma_{2}$, then Proposition 5(c) implies there exists a critical level of affective polarization $\alpha_{3} \in (0, \hat{\alpha})$ such that $\Delta Q^{\ast}(\hat{\alpha}, 0) = 0$. \\
	
\noindent {\it Case 2: $P =A$ and $P'=F$ such that $I_{A}^{0} > I_{F}^{0}$}. Using Lemma 2 and $\mathcal{Q}^{\ast}(\alpha = 0)$, we get
\begin{align}\notag
\Delta Q^{\ast}(\hat{\alpha}, 0) ~=~\sum\limits_{i \in A} I^0_{iA} ~ \Big(u(m_{i}^{\ast})\big|_{\alpha \rightarrow \hat{\alpha}} ~-~ u(m_{i}^{\ast})\big|_{\alpha = 0} ~\Big) ~-~ \sum\limits_{i \in F} I^{0}_{iF} ~ \Big(~ u(m_{i}^{\ast})\big|_{\alpha = 0} ~+~ \sigma ~ \Big) \\
~=~ \sum\limits_{i \in A} I^{0}_{iA} ~ \Big(u(m_{i}^{\ast})\big|_{\alpha \rightarrow \hat{\alpha}} ~-~ u(m_{i}^{\ast})\big|_{\alpha = 0} ~\Big) ~-~ \sum\limits_{i \in F} I^{0}_{iF} ~ \Big(~ u(m_{i}^{\ast})\big|_{\alpha = 0} \Big) ~-~ \sigma \cdot I_{F}^{0}
\end{align}
	
\noindent The above equation suggests $\Delta Q^{\ast}(\hat{\alpha}, 0)$ will be strictly positive if 
$$ \sigma ~<~ \sum\limits_{i \in A} \big( \frac{I_{iA}^{0}}{I_{F}^{0}} \big) \cdot \Big(u(m_{i}^{\ast})\big|_{\alpha \rightarrow \hat{\alpha}} ~-~ u(m_{i}^{\ast})\big|_{\alpha = 0} ~\Big) ~-~ \sum\limits_{i \in A} \big( \frac{I_{iA}^{0}}{I_{F}^{0}} \big) \cdot \big( u(m_{i}^{\ast})\big|_{\alpha = 0} \big)\equiv \sigma_{3}$$
	
\noindent If $\sigma < \sigma_{3}$, then Proposition 5(c) implies there exists a critical level of affective polarization $\alpha_{4} \in (0, \hat{\alpha} )$ such that $\Delta Q^{\ast}(\hat{\alpha}, 0) ~=~ 0$.
	
\end{document}